\documentstyle[12pt,epsfig]{article}
\def\beq{\begin{equation}}
\def\eeq{\end{equation}}
\def\bea{\begin{eqnarray}}
\def\eea{\end{eqnarray}}
\def\bq{\begin{quote}}
\def\eq{\end{quote}}
\newcommand{\ber}{\begin{eqnarray}}
\newcommand{\eer}{\end{eqnarray}}

\def\gappeq{\mathrel{\rlap {\raise.5ex\hbox{$>$}}
{\lower.5ex\hbox{$\sim$}}}}
\def\lappeq{\mathrel{\rlap{\raise.5ex\hbox{$<$}}
{\lower.5ex\hbox{$\sim$}}}}
\def\Toprel#1\over#2{\mathrel{\mathop{#2}\limits^{#1}}}

\begin{document}
\pagestyle{empty}
\begin{flushright}
{CERN-TH/2001-342}\\
\end{flushright}
\vspace*{5mm}
\begin{center}
{\bf The Statistical Model of Hadrogenesis in A--A collisions\\
from AGS to SPS and RHIC}$^{*)}$\\
\vspace*{1cm}
{\bf Krzysztof Redlich$^{a,+)}$ }\\
 $^a$
Theoretical Physics Division, CERN, CH-1211 Geneva 23, Switzerland  \\
 \vspace*{2cm} {\bf ABSTRACT} \\
  \end{center} \vspace*{1mm}
\noindent We discuss   experimental data on particle yields and
particle spectra obtained  in heavy ion collisions in  a very
broad energy range  from  SIS/GSI ($\sqrt s\simeq 2$ GeV) through
AGS/BNL ($\sqrt s\simeq ~~5$ GeV) up to SPS/CERN  ($\sqrt s\simeq
20$ GeV) and RHIC/BNL ($\sqrt s\simeq 130$) GeV. We argue that in
this broad energy range  hadronic yields and their ratios resemble
a thermal equilibrium population along a  unified  freeze--out
curve determined by the condition of fixed energy/particle $\simeq
1$ GeV. At RHIC and top SPS,  thermal parameters are consistent
within error with the critical conditions required for
deconfinement. This, together with the particular distribution of
strangeness within a collision fireball, could indicate that
chemical equilibrium is a direct consequence of parton to hadron
transition, which populates a state of maximum entropy. At lower
energies equilibration in A--A collisions  should appear through
hadronic interactions and rescatterings.

\vspace*{4.5cm} \noindent \rule[.1in]{16.5cm}{.002in}

\noindent $^{+)}$ Permanent address: Institute of Theoretical
Physics, University of Wroc\l aw, PL-50204 Wroc\l aw, Poland:
redlich@rose.ift.uni.wroc.pl \vspace*{0.5cm}\\
$^*)$ Plenary talk given at: International Nuclear Physics
Conference, INPC2001, Berkeley, USA,  August 2001.

\begin{flushleft} CERN-TH/2001-342\\
December 2001
\end{flushleft}
\vfill\eject

\setcounter{page}{1} \pagestyle{plain}

\section{Introduction}

The ultimate goal of ultrarelativistic nucleus--nucleus collisions
is to study the properties of strongly interacting matter  under
extreme conditions of high energy density \cite{satz}. Quantum
Chromodynamics (QCD) predicts that strongly interacting matter
undergoes a phase transition from a state of hadronic constituents
to a plasma  of unbounded quarks and gluons (QGP)
\cite{shuryak,hatsuda}. By colliding heavy ions at
ultrarelativistic energies, one expects to create hadronic matter
under that conditions  are sufficient for deconfinement
\cite{satz,stock,stachel,uh,jr}. Thus, of particular relevance was
finding experimental probes to check if the produced medium in its
early stage was indeed in the QGP phase. Different probes have
been theoretically proposed and studied  in terms of  SPS/CERN and
most recently  RHIC/BNL experiments. The most promising signals of
QGP were related with particular properties of photons
\cite{alam,dinesh}, dileptons \cite{alam,rapp} and hadron spectra
\cite{stachel,uh,wideman}. The photon rate was expected to be
enhanced if the QGP were  formed in the initial state. The
invariant mass distribution of dileptons should be modified by
in-medium effects related with chiral symmetry restoration
\cite{rapp,gery}. The suppression of charmonium production was
argued to be a consequence of the collective effects in a
deconfined medium \cite{satz,matsui}.

Hadron multiplicities  and their correlations are observables
which can provide  information on the nature, composition, and
size of the medium from which they are originating. Of particular
interest  is the extent to which the measured particle yields are
showing equilibration. The appearance of the QGP, that is a
partonic medium being at (or close to) local thermal equilibrium,
and its subsequent hadronization during phase transition should in
general drive hadronic constituents towards chemical equilibrium
\cite{stock,stachel,uh,knol}. Consequently, a high level of
chemical saturation, particularly for strange particles
\cite{rafelski}, could be related with the deconfined phase
created at the early stage of heavy ion collisions.

The level of equilibrium of secondaries in heavy ion collisions
was tested by analysing the particle abundances
\cite{stachel,uh,Bra99,CLK,bra95,becattini2} or their momentum
spectra \cite{wideman,tomasik,vesa}. In the {\rm first} case one
establishes the chemical composition of the system, while in the
second  case additional information on dynamical evolution and
collective flow can be extracted.

In this article we will discuss and analyse  the experimental data
on hadronic abundances obtained in ultrarelativistic heavy ion
collisions, in a very broad energy range starting from RHIC/BNL
($\sqrt s=130$ A GeV),           SPS/CERN ($\sqrt s\simeq 20$ A
GeV) up to AGS/BNL ($\sqrt s\simeq 5$ A GeV) and SIS/GSI ($\sqrt
s\simeq 2$ A GeV) to test  equilibration. We argue  that the
statistical approach  provides a very satisfactory description of
experimental data covering a wide energy  range from SIS up to
RHIC. We discuss the unified description of particle chemical
freeze--out and the excitation function of different particle
species. Introducing, in addition to thermal,  also the transverse
collective motion, the systematics of thermal freeze--out is also
presented.

\section{Initial conditions in A--A  collisions and deconfinement}
In ultrarelativistic heavy ion collisions, the knowledge of the
critical energy density $\epsilon_c$ required for deconfinement is
of particular importance  as well as the equation of state (EoS)
of strongly interacting matter. The value of $\epsilon_c$ is
needed to establish the necessary initial conditions in heavy ion
collisions to possibly create the QGP, whereas EoS is required as
an input to describe  the space-time evolution of the collision
fireball.

Both of these these pieces of information can be obtained today
from first principal calculations by formulating QCD on the
lattice and performing  numerical Monte-Carlo simulations. In
Fig.~1 we show the most recent results of lattice gauge theory
(LGT) for energy density and pressure \cite{karsch}. These results
have been obtained in LGT for different numbers of dynamical
fermions. The energy density is seen in Fig.~1 to exhibit a
typical behaviour in a system with a phase transition: an abrupt
change in the very narrow temperature range.\footnote{We have to
point out, however, that in the strictly statistical physics
sense, a phase transition can only appear in the limit of massless
quarks.} The corresponding pressure  shows a smooth change with
temperature. In the region below $T_c$ the basic constituents of
QCD, quarks and gluons, are confined within their hadrons and here
the EoS is well parametrized by the hadron resonance gas. Above
$T_c$ the system appears in the QGP phase where quarks and gluons
can penetrate distances that substantially exceed    a typical
size of hadrons. The most recent results of improved perturbative
expansion of thermodynamical potential in the continuum QCD
\cite{blaizot} are showing that at some distance above $T_c$ the
EoS of QGP can be well described by a gas of massive
quasi-particles whose   mass is temperature dependent. In the near
vicinity of $T_c$ the relevant degrees of freedom were argued to
be described by the Polyakov loops \cite{rob}.

Lattice Gauge Theory  predicts that in two--flavour QCD the
critical temperature $T_c=173\pm ~8$MeV and the  corresponding
critical energy density $\epsilon_c=0.6\pm 0.3$ GeV/fm$^3$ are
required for chiral phase transition \cite{karsch}. The value of
$\epsilon_c$ is relatively low and quantitatively corresponds to
the energy density inside the nucleon.
\begin{figure}[htb]
\begin{minipage}[t]{80mm}
\includegraphics[width=16.5pc, height=14.3pc]{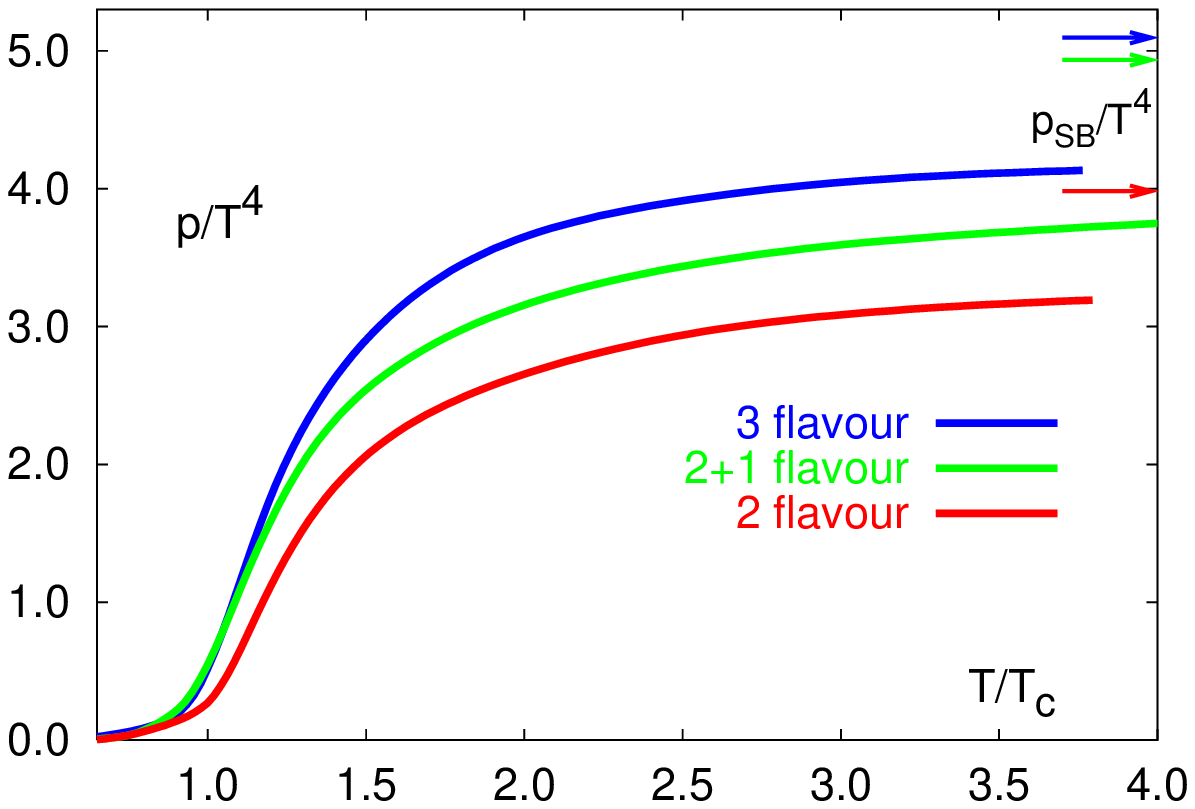}\\
\end{minipage}
%
%
\begin{minipage}[t]{80mm}
\includegraphics[width=17.5pc, height=14.2pc]{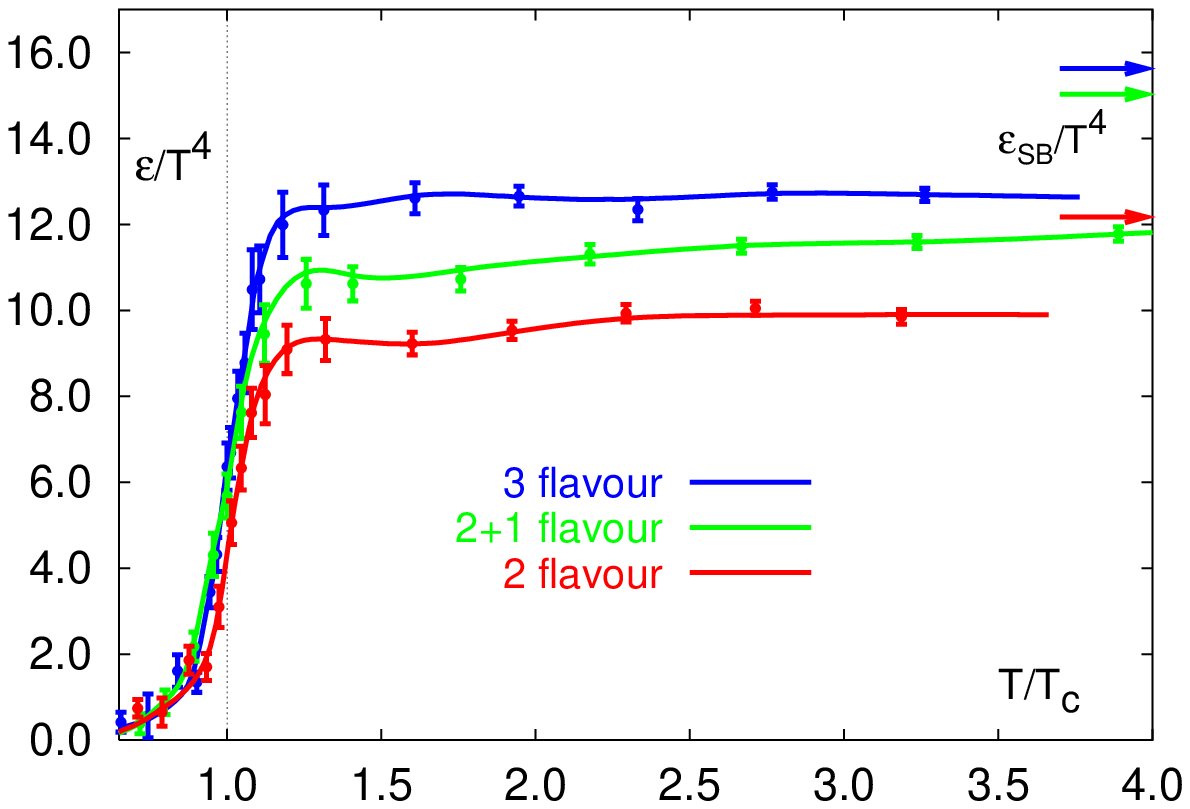}\\
\end{minipage}
\begin{minipage}[t]{169mm}
\caption{Pressure $P$ and energy density  normalized to
temperature in fourth power,
 versus temperature normalized to its critical value. The calculations were done within LGT for different
 numbers of flavours \cite{karsch}. The values of the corresponding ideal gas results  are indicated by arrows. }
\end{minipage}
\end{figure}

The initial energy density reached in  heavy ion collisions can be
estimated within the Bjorken model \cite{bjorken}. From the
rapidity distribution of protons and  their transverse energy
$E_T$ measured in nucleus--nucleus  collisions  the initial energy
density $\epsilon_0$ is determined from

\begin{equation}
\epsilon_0(\tau_0)={1\over {\pi R^2}}{1\over \tau_0}{{dE_T}\over
{dy}}.\label{eq1}
\end{equation}
where the  initially produced collision fireball is considered as
a cylinder of length $\tau_0dy$ and transverse size $R\sim
A^{1/3}$. Inserting for $\pi R^2$ the overlap area of colliding Pb
nuclei together with initial time an $\tau_0\simeq 1$ fm, and
using an average transverse energy at midrapidity measured at the
SPS ($\sqrt s=17.3$ GeV) to be 400 GeV \cite{400},   one obtains
\begin{equation}
\epsilon_0^{SPS}(\tau_0\simeq 1~ {\rm fm})\simeq 3.5\pm 0.5 ~{\rm
GeV/fm^3}.\label{eq2}
\end{equation}

Increasing the collision energy to $\sqrt s=130$ A GeV for Au--Au
at RHIC and keeping the same initial thermalization time as at the
SPS, one would expect an increase of $\epsilon_0$ by only 50--60
$\%$. However, at RHIC the thermalization is argued, within
saturation models, to appear at much shorter time. The basic
concept of saturation models is a conjecture  that there is some
transverse momentum scale $p_{\rm sat}$ where the gluon and quark
phase space density saturates \cite{sat}.  For isentropic
expansion of the collision fireball, one can relate the transverse
energy at $p_{\rm sat}$ with the one measured in nucleus--nucleus
collisions in the final state. The saturation scale also fixes the
time scale $\tau_0=1/p_{\rm sat}$. Taking the value of $p_{\rm
sat}$ predicted in \cite{kari} for RHIC $p_{\rm sat}\simeq 1.13$
GeV, one gets $\tau_0\simeq 0.2$ fm and the corresponding initial
energy density $\epsilon_0\simeq 98$ GeV/fm$^3$.
The estimate of  $\epsilon_0$, however, strongly depends on the
value of $p_{\rm sat}$, which is model--dependent.  The
McLerran--Vanugopalan model \cite{larry}, for instance,  predicts
the value of $\epsilon_0^{RHIC}\sim 20$ GeV/fm$^3$,  which agrees
with the predictions of  \cite{nardi}. At the SPS the saturation
model described in \cite{kari} leads to $\epsilon_0^{SPS}\sim 16$
GeV/fm$^3$,  a  much higher value than given in Eq.~(2). It is
thus clear that  there are large uncertainties on the value of the
initial energy density reached in ultrarelativistic heavy ion
collisions. In Pb--Pb collisions at the SPS according to the
models one gets $2.5 ~{\rm GeV/fm^3}< \epsilon_0^{SPS}< 16~ {\rm
GeV/fm^3}$ whereas in Au--Au collisions at RHIC,~  $20 {\rm
~GeV/fm^3}< \epsilon_0^{SPS}< 100~ {\rm GeV/fm^3}$.

The dominant component of the partonic medium produced in
ultrarelativistic heavy ion collisions at RHIC and even at the SPS
is gluons. The energy density of gluons in thermal equilibrium
scales with the fourth power of the temperature $\epsilon=gT^4$,
where $g$ denotes the number of degrees of freedom. For an ideal
gas, $g=16\pi^2/30$; in an interacting system, the effective
number of degrees of freedom $g$ is smaller. The results of LGT
shown in Fig.~1 indicate   deviations from the Boltzmann limit by
20--25 $\%$. Relating the thermal energy density with the initial
energy density discussed above, one can make an estimate of the
initial temperature reached in heavy ion collisions. For the
 SPS this gives a temperature in the range $200~  {\rm MeV} < T <
330$ MeV and at RHIC $400 ~ {\rm MeV}< T<  600$ MeV.

\begin{figure}[htb]
 {\hskip 2.1cm
\includegraphics[width=30.5pc, height=15.9pc]{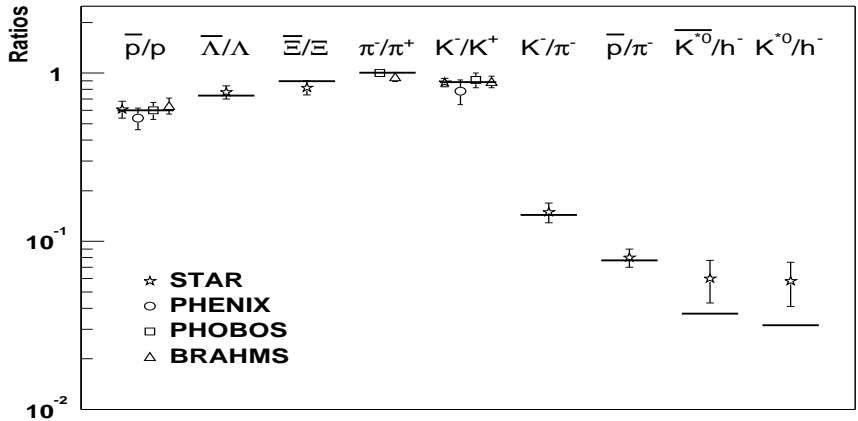}}\\
\vskip -0.5cm { \caption{Comparison of the  experimental data on
different particle multiplicity ratios obtained  at RHIC at
$(\sqrt s=130)$ GeV  with   thermal model calculations for $T=175$
MeV and $\mu_B=51$ MeV.
 } \label{rhic}}
\end{figure}

Comparing the  initial energy density  expected in heavy ion
collisions with   LGT results, it is clear that  the initial
energy density, at RHIC, exceeds by far the critical value.
Thus, the necessary conditions to create the partonic medium in a
deconfined phase are reached  at RHIC as well as  at the  top SPS.
Large energy density is, however,  still not sufficient to create
a QGP. The distribution of initially produced gluons is very far
from being thermal, thus the system needs enough time to
equilibrate. Recently,  it was rigorously shown \cite{rolf} in the
framework of perturbative QCD and kinetic theory  that the
equilibration of partons  happens indeed at the LHC and most
likely at RHIC. Previous microscopic study within the Parton
Cascade Model has also suggested that thermalization can be
reached at lower SPS energy \cite{pc}. Here, however, it is not
clear if models  inspired by the perturbative QCD are indeed
applicable at this relatively low collision energy.

Admitting  QGP formation in the initial state in heavy ion
collisions  one could  expect that the thermal nature of the
partonic medium could be preserved during hadronization.
Consequently,  the particle yields measured in the final state
should resemble the thermal equilibrium population. In the
following, we present the most recent results related with the
question of equilibration of secondaries in heavy ion collisions
and discuss its possible relation with deconfinement.

\section{Statistical model and particle multiplicity}

The basic quantity in the statistical model description of thermal
properties of hadronic matter is the partition function $Z(T,V)$.
In the Grand Canonical (GC) ensemble,

\begin{equation}
Z^{GC}(T,V,\mu_Q)\equiv {\rm Tr}[e^{-\beta (H -\sum_i\mu_{Q_i}
Q_i)}],
\end{equation}
where  $H$ is the hamiltonian of the system, $Q_i$ are the
conserved charges and  $\mu_{Q_i}$  is the chemical potentials
that guarantees that the charge $Q_i$ is conserved on the average
in the whole system. Finally $\beta =1/T$ is the inverse
temperature.

In the strongly interacting medium, one includes the conservation
of electric charge, baryon number and strangeness. Thus, the
partition function depends in general on five parameters. However,
only three are independent,  since the isospin asymmetry in the
initial state fixes the charge chemical potential and strangeness
neutrality conditions eliminate the strange chemical potential. On
the level of particle multiplicity ratios derived from the
partition function, we are  left  with only  temperature $T$ and
baryon chemical potential $\mu_B$ as independent parameters.
\begin{figure}[htb]
\begin{minipage}[t]{80mm}
\includegraphics[width=14.5pc, height=17.2pc]{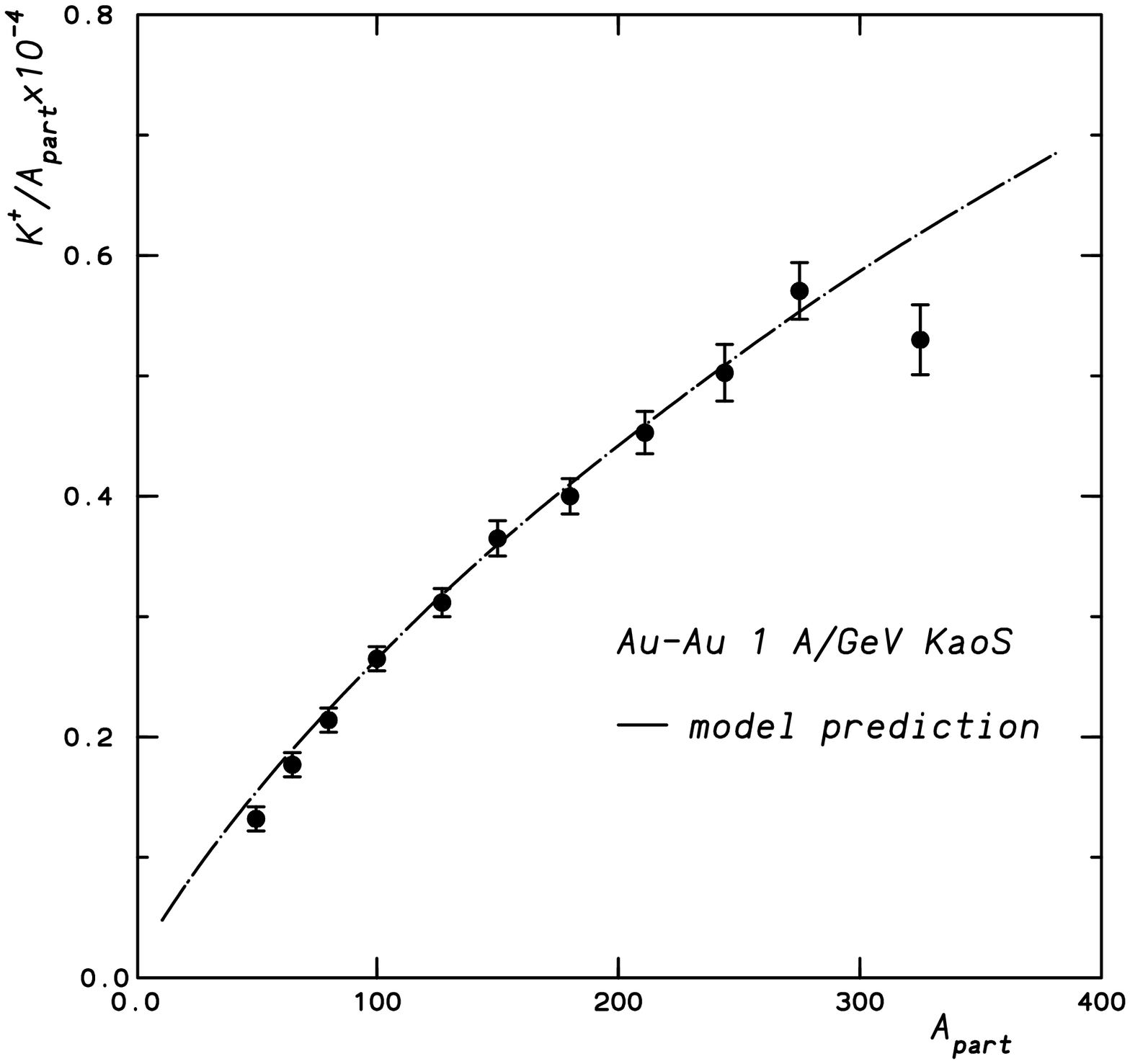}\\
\end{minipage}
%
%
\begin{minipage}[t]{75mm}
\includegraphics[width=22.5pc, height=18.2pc]{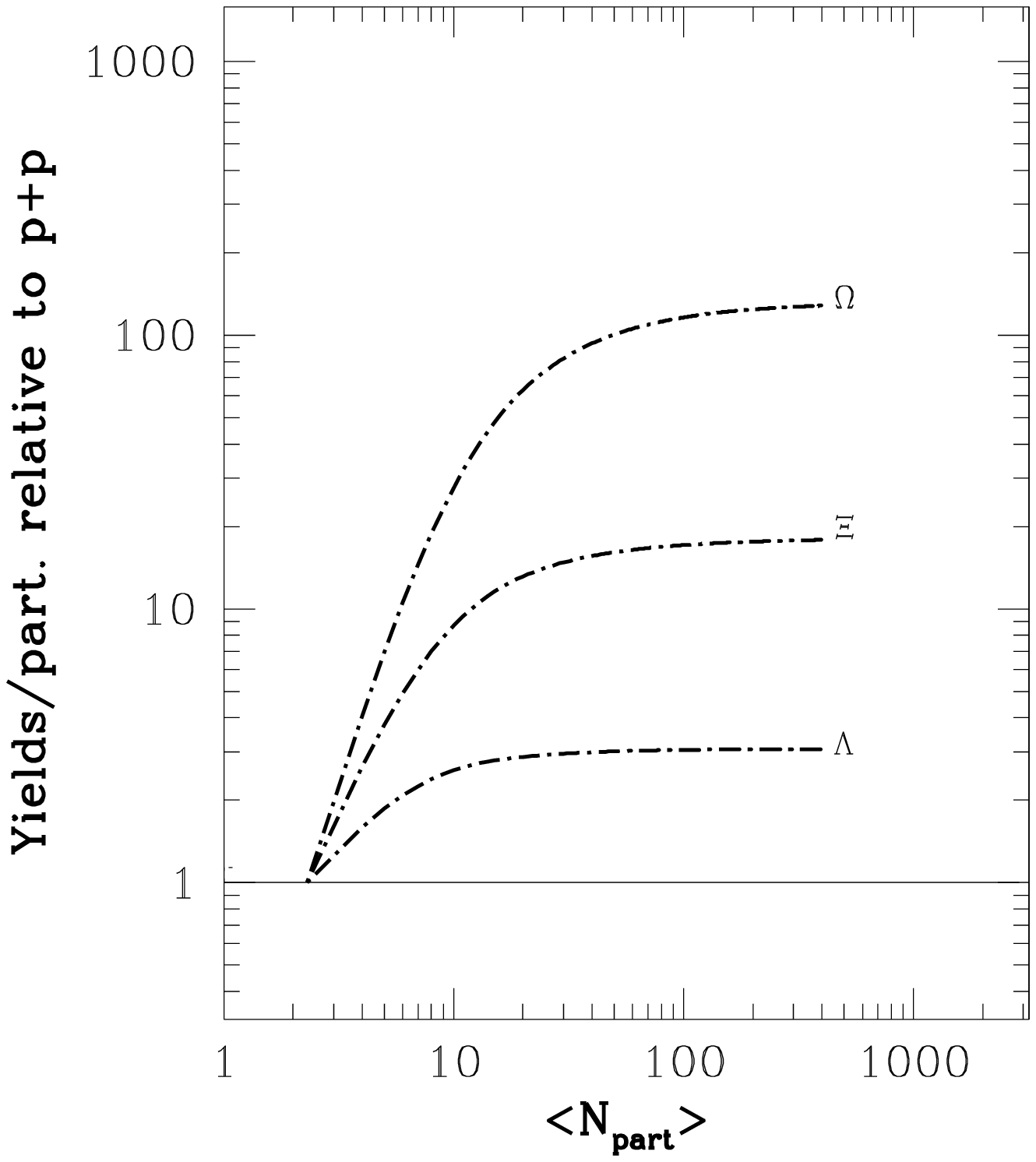}\\
\end{minipage}
\begin{minipage}[t]{169mm}
 {\vskip -0.7 true cm
 \caption{On the left, the ratio of kaon to pion measured in
Au--Au collisions at 1 A GeV \cite{data}; the broken line
represents the statistical model results \cite{Cle99}. The
rights--hand figure shows statistical model predictions for
yield/participants in A--A collisions at 40 A GeV  normalized to
the corresponding value in pp collisions.}}
\end{minipage}
\end{figure}

The Hamiltonian is usually described by the hadron resonance gas,
which contains the contributions from  all mesons with masses
below 1.6 GeV and baryons with masses below 2 GeV. In this mass
range the hadronic spectrum is well established and the decay
properties of resonances are  known. This mass cut in the
contribution to partition function limits, however, the maximal
temperature to $T_{\rm max}<190$ MeV,  up to which the model
predictions could be trustworthy. For higher temperatures the
contributions of heavier resonances are not negligible.

In the high  density regime the repulsive interactions of hadrons
should also be included in the partition function. To incorporate
the repulsion on short distances between hadrons one usually uses
a hard core description by implementing the excluded volume
corrections. In the thermodynamically consistent approach
\cite{tc} these corrections lead to a shift of baryon chemical
potential. Finally, the widths of resonances and their decay into
lighter particles have to be included in the statistical model
when calculating particle multiplicities \cite{bra95,Cle99}.

The statistical model, described above, was applied to Pb--Pb
collisions at top SPS energy \cite{CLK,bra95,becattini2}. The
model was compared with almost all experimental data obtained by
NA44, NA49 and WA97 Collaboration. Hadron multiplicities ranging
from pion to omega and their ratios were used to verify if there
is a set of thermal parameters $(T,\mu_B)$ that simultaneously
reproduces all measured yields. A detailed analysis has shown
\cite{bra95} that choosing a temperature $T=168\pm 4$ MeV and a
baryon chemical potential $\mu_B=266\pm 8$ MeV, the statistical
model with only two parameters can indeed describe seventeen
different particle multiplicity ratios within an accuracy of one
to two standard deviations. One could thus conclude that, with
respect to the statistical operator formulated for equilibrium
hadron resonance gas, the experimental data at the SPS are showing
a high level of chemical equilibration. The natural question
arising here would be to what extent this statistical operator
provides a unique description of the data.  This question was
addressed in the literature and two distinct models have been
examined \cite{jr,cise}. In \cite{cise} the authors analysed the
possible influence of in-medium effects  on the chemical
equilibrium description of particle yields at the SPS. In
\cite{jr} the non-equilibrium scenario of explosive hadronization
of a QGP fireball was proposed. Both these models and particularly
\cite{jr} are showing satisfactory agreement with SPS data,
however with larger deviations for multistrange particles. In our
discussion we concentrate on equilibrium description of particle
production since only this approach, as  will be demonstrated,
provides the systematic agreement with almost all heavy ion data
from SIS to RHIC.

The chemical freeze--out temperature,  found  from a thermal
analysis \cite{bra95,becattini2} of experimental data in Pb--Pb
collisions at the SPS is remarkably consistent, within errors,
with the critical temperature $T_c\simeq 173\pm 8$ MeV obtained
from lattice Monte-Carlo simulations of QCD at vanishing baryon
density \cite{karsch}. Thus, the observed hadrons  seem to be
originating from a deconfined medium and the chemical composition
of the system is most likely to be established during
hadronization \cite{stock,stachel,uh}. The observed coincidence of
chemical and critical conditions in the QCD medium, if indeed
valid, should be  seen also in heavy ion collisions at higher
collision energies, in particular at RHIC.

\begin{figure}[htb]
 {\hskip -.1cm
\includegraphics[width=35.5pc, height=13.9pc]{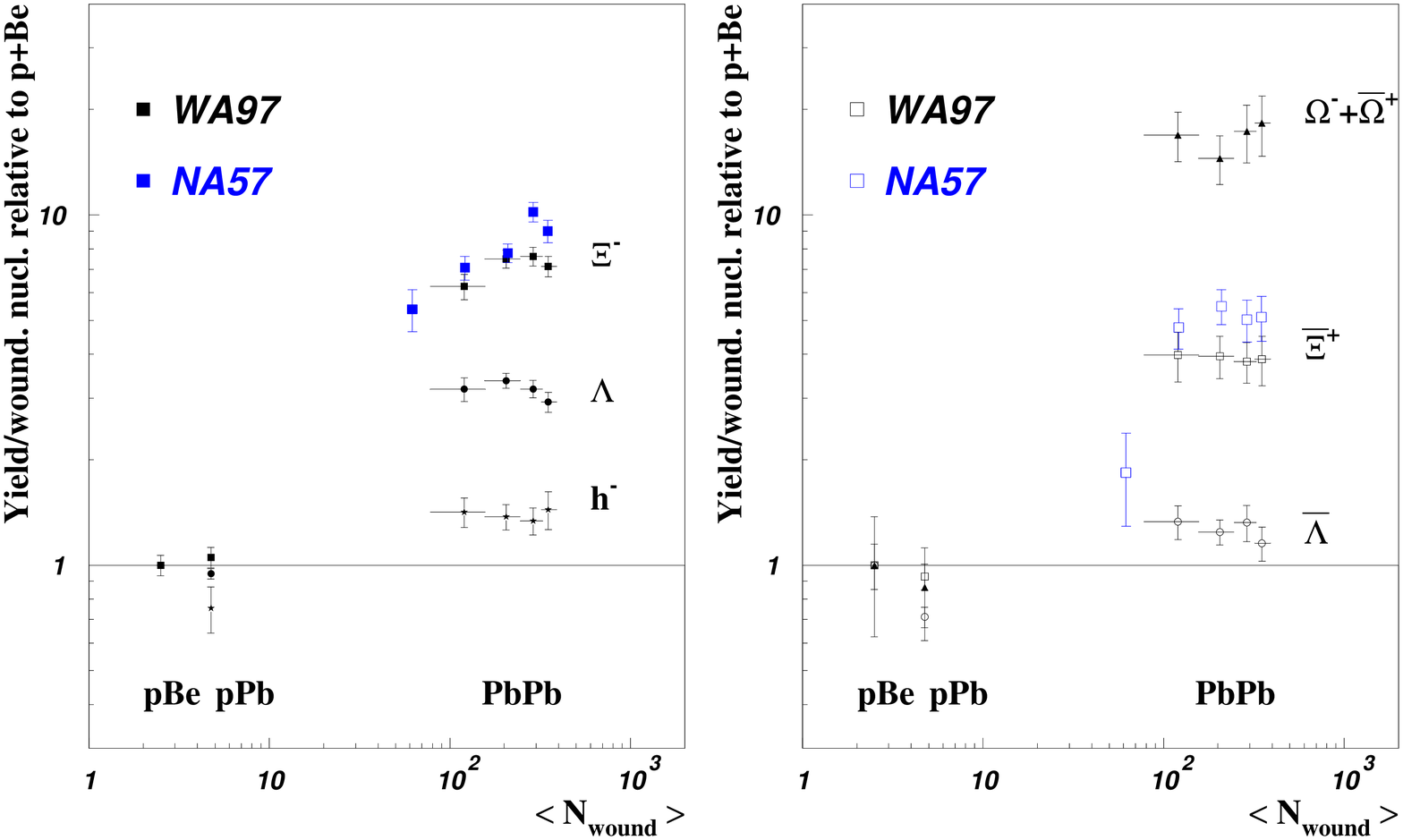}}\\
{ \caption {Particle yields per participant in Pb+Pb relative to
p+Be and p+Pb collisions centrality dependence. The data are from
WA97 \cite{wa97} and NA57 \cite{na57} Collaborations.}}
\end{figure}

The equilibrium statistical model was recently  applied to Au--Au
collisions at $\sqrt s=$~130 GeV at RHIC \cite{damian}.  The
results of the STAR,
PHENIX, PHOBOS and BRAHMS  Collaboration for different particle
multiplicity ratios have been used to test chemical equilibration
at RHIC.  In Fig.~2 we show the comparison  of the thermal model
with experimental data. One sees  that the overall agreement is
very good. Most of the data are reproduced by the model within the
experimental errors. The largest deviations are seen in the ratios
of ${\bar K^{*0}}/h^-$ and $K^{*0}/h^-$ but they are still on the
level of one standard deviation.\footnote{ In the model
calculations, the $h^-$ was  considered   as  the  total number of
negatively charged particles in the thermal fireball} In Au--Au
collisions at $\sqrt s$=~130~GeV the chemical freeze--out appears
at $T= 175\pm 7$ MeV and $\mu_B = 51\pm 6$ MeV. The resulting
temperature is only slightly higher than that previously found at
the SPS where for Pb-Pb collisions $T=168\pm 5$MeV.  This
relatively moderate increase of temperature could be expected
since, in the limit of vanishing baryon density, the temperature
should not exceed the critical value required for deconfinement.
The substantial decrease of the baryon chemical potential from
$\mu_B\simeq 270$ MeV at the SPS to $\mu_B\simeq 50 $ MeV  at RHIC
shows that, at midrapidity, we are dealing with a low net baryon
density  medium.

The results for particle yields and their ratios at the SPS and
RHIC  shows the statistical order. Chemical equilibration of
secondaries after hadronization is rather excluded by kinetics
\cite{rafelski,kinetics}. Thus, the equilibrium population of
hadrons would be most likely to appear since it was
pre--established in the QGP phase. In the following we argue,
however, that equilibration of secondaries is not a unique signal
for deconfinement, as it is also there at lower energies, where
the initial conditions exclude QGP formation. To test
equilibration in low energy nucleus-nucleus collisions, one needs,
however, to change the statistical operator from a GC to a
canonical C ensemble with respect to  strangeness conservation.

\section{Equilibrium limit of rarely produced  particles}

The conservation of quantum numbers related with U(1) internal
symmetry in statistical models can be described in the GC ensemble
only if the number of produced  particles per event carrying
corresponding quantum number  is much larger than 1. In the
opposite limit of rare particle production \cite{Hag71,ko}, U(1)
charge  conservation must be implemented locally on an
event-by-event basis, i.e., a canonical C ensemble of conservation
laws must be used. The C ensemble is relevant in the statistical
description of particle production in low energy heavy ion
\cite{Cle99}, or high energy hadron--hadron or $e^+e^-$ reactions
\cite{b1} as well as in peripheral heavy ion collisions
\cite{hamieh}.
\begin{figure}[htb]
\begin{minipage}[t]{90mm}
\vskip 2.5cm {\hskip 0.5cm
\includegraphics[width=17.5pc, height=16.5pc]{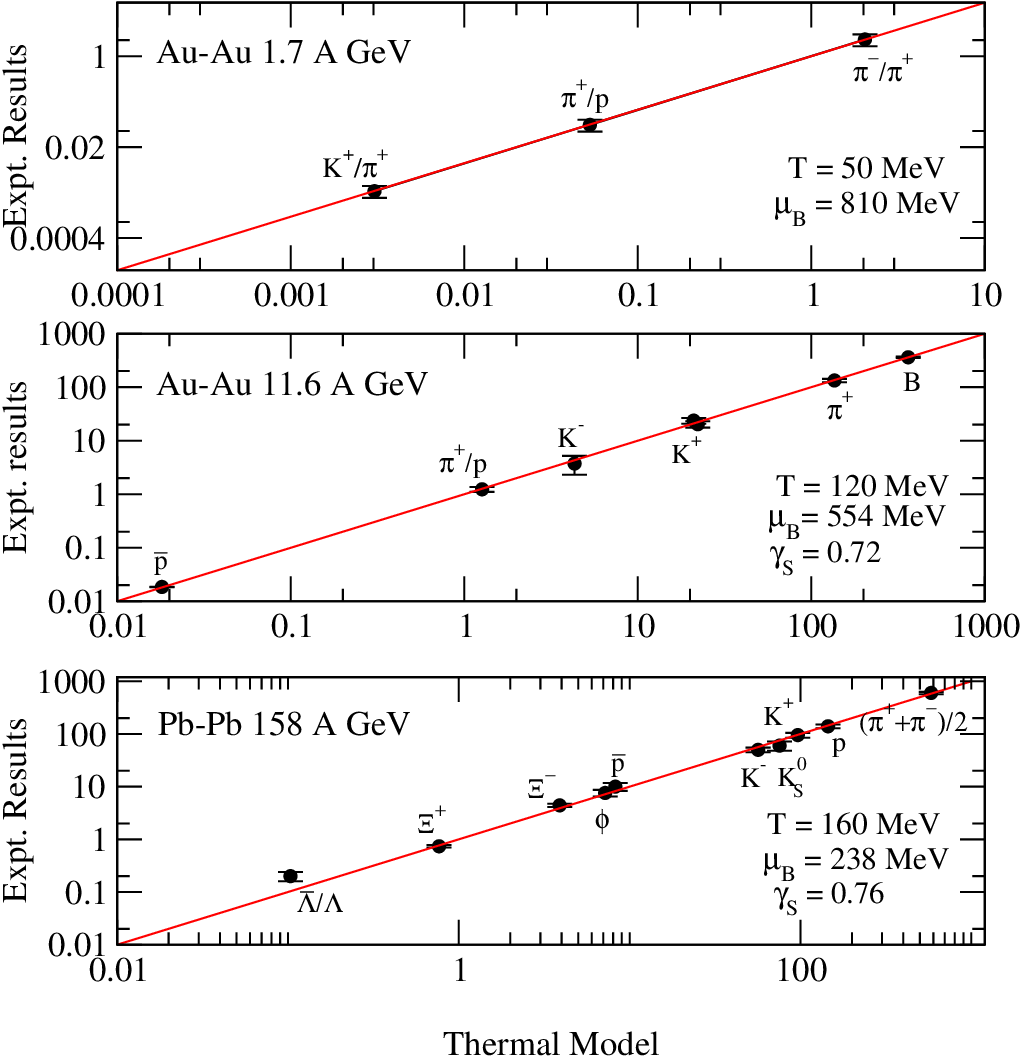}}\\
\end{minipage}
\hspace{\fill}
\begin{minipage}[t]{58mm}
\hskip -7.0cm {\vskip 2.5cm
 {\vskip 0.5cm  {{\bf FIGURE 4b}.\noindent

Comparisons of the experimental data (vertical scale) for
different particle multiplicity ratios obtained in Au--Au and
Pb--Pb collisions at SIS, AGS and SPS  with the statistical model
predictions \cite{becattini2}.
 }}}
\end{minipage}
\end{figure}


The exact conservation of quantum numbers, that is the canonical
approach, is known to  severely reduce the thermal phase--space
available for particle production \cite{Hag71}. Consequently, the
chemical equilibrium limit of rarely produced particles is changed
and it is different from the one obtained in the asymptotic GC
limit. In order to illustrate the above change, let us consider
the kinetics for the time evolution and equilibration of rarely
produced particles by considering a simple example of $K^+K^-$
pair production and equilibration in the environment of thermal
pions contained in volume $V$ at temperature $T$. The production
of kaons is due to the  binary process $\pi^+\pi^-\to K^+K^-$.

In the standard formulation  \cite{koch}, the rate equation for
this binary process is described by the following population
equation:
\begin{eqnarray}
\frac{d\langle N_K\rangle }{d\tau}={G\over V} \langle
N_{\pi^+}\rangle  \langle N_{\pi^-}\rangle  - {L\over V} \langle
N_{K^+}\rangle\langle  N_{K^-}\rangle , \label{normal1}
\end{eqnarray}
where $G \equiv \langle \sigma_G v \rangle$ and $L \equiv \langle
\sigma_L v \rangle$ give the momentum-averaged cross sections for
the gain  $\pi^+ \pi^- \rightarrow K^+ K^-$ and the loss  $K^+ K^-
\rightarrow \pi^+ \pi^-$ process, respectively, and $\langle
N_K\rangle $ represents the total number of produced kaons.

In the above equation it is explicitly assumed that $K^+$ and
$K^-$ are uncorrelated. To include  possible correlations between
the production of $K^+$ and $K^-$ \cite{ko}, let us define
$P_{i,j}$ as the probability to find $i$ number of  $K^+$  and $j$
number of $K^-$ in an event. We also denote by $P_i$ the
probability to find $i$ number of  $K$ in an event. The average
number of  $K$ per event is defined as:
$\langle N_K \rangle =\sum_{i=0}^{\infty} iP_i.$

We can now write the following general rate equation for the
average kaon multiplicities:
\begin{eqnarray}
\frac{d\langle N_{K} \rangle }{d\tau}= {G\over V} \langle
N_{\pi^+} \rangle \langle N_{\pi^-} \rangle - \frac {L}{V}
\sum_{i,j} ij P_{i,j}. \label{general}
\end{eqnarray}

Owing to the local conservation of quantum numbers, we have:
\begin{eqnarray}
P_{i,j}&=& P_i ~\delta_{ij}, \nonumber\\
\sum_{i,j} ij P_{i,j} &=& \sum_i i^2 P_i \equiv \langle N^2
\rangle =\langle N \rangle ^2+\langle \delta N^2 \rangle,
\label{5}
\end{eqnarray}
where $\langle \delta N^2 \rangle$ represents the event-by-event
fluctuation of the number of $K^+ K^-$ pairs. Note that we always
consider abundant $\pi^+$ and $\pi^-$  so that we can neglect the
number fluctuation of these particles and the change of their
multiplicities due to the considered processes.
\begin{figure}[htb]
 {\hskip 2.1cm
\includegraphics[width=27.5pc, height=22.9pc]{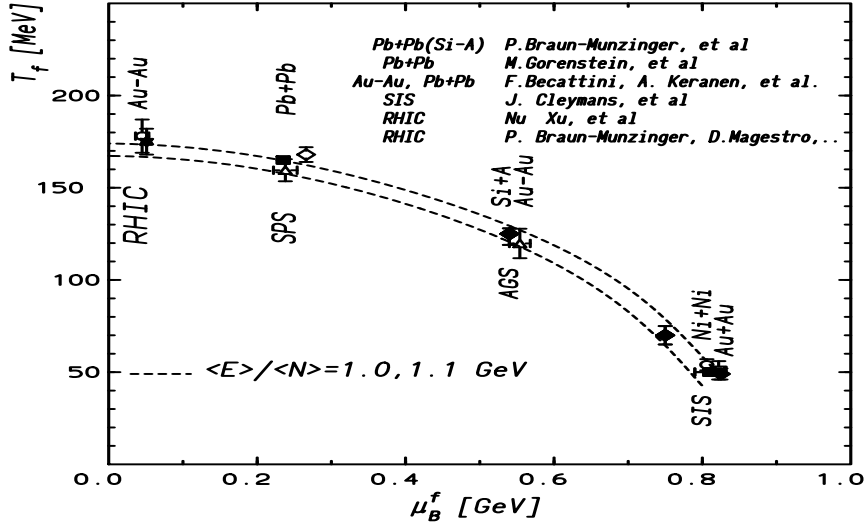}}\\
{\vskip -1.8cm \caption{Compilation of chemical freeze--out
parameters from SIS to RHIC. The broken line represents a
phenomenological condition of chemical freeze--out of    fixed
energy/particle $\simeq 1$ GeV \cite{CLK}. }\label{carrer}}
\end{figure}

Following  Eqs.~(\ref{general}) and (\ref{5}) the general rate
equation for the average number of $K^+K^-$ pairs can be written
as:
\begin{eqnarray}
\frac{d\langle N_K \rangle }{d\tau}={G\over V} \langle N_{\pi^+}
\rangle \langle N_{\pi^-} \rangle - \frac {L}{V} \langle N_K^2
\rangle . \label{general2}
\end{eqnarray}
For abundant production of $K^+ K^-$ pairs where $\langle N_K
\rangle \gg 1$,
$\langle N_K^2 \rangle \approx \langle N_K\rangle ^2,$
and Eq.~(\ref{general2}) obviously reduces to the standard form:
\begin{eqnarray}
\frac{d\langle N_K\rangle }{d\tau}\approx \frac{G}{V} \langle
N_{\pi^+} \rangle \langle N_{\pi^-} \rangle - \frac{L}{V} \langle
N_{K^+}\rangle\langle N_{K^-}\rangle . \label{normal2}
\end{eqnarray}
However, for rare production of $K^+ K^-$ pairs where $\langle
N_K\rangle\ll \!1$, the rate equations (\ref{normal1}) and
(\ref{normal2}) are no longer valid. We have instead
$\langle N_K^2 \rangle \approx \langle N_K\rangle ,$
which reduces Eq.~(\ref{general2}) to the form \cite{ko}:
\begin{eqnarray}
\frac{d\langle N_K\rangle }{d\tau}\approx {G\over V}
\langle N_{\pi^+} \rangle \langle N_{\pi^-} \rangle
- \frac{L}{V} \langle N_K \rangle . \label{canonical}
\end{eqnarray}
Thus, the limit where $\langle N_K \rangle \ll 1$, the absorption
term depends on the pair number only linearly, instead of
quadratically for the limit of $\langle N_K \rangle \gg 1$.  It is
thus clear, that   the time evolutions and equilibrium values for
kaon multiplicities are obviously different in the above limiting
situations.

In the limit of large $\langle N_K\rangle$ the equilibrium value
for the number of $K^+K^-$ pairs, which coincides with the
multiplicity of $K^+$ and $K^-$, is obtained from
Eq.~(\ref{normal2}) as,

\begin{eqnarray}
\langle N_K\rangle_{\rm eq}^{\rm GC}= { {V}\over {2\pi^2}} m_K^2 T
K_2 (M_K/T), \label{15}
\end{eqnarray}
 it is thus described by the  GC result with vanishing chemical
potential, in our example, because of the  strangeness  neutrality
condition.

In the opposite limit, where $\langle N_K \rangle \ll 1$, the time
evolution of pion multiplicity is described by
Eq.~(\ref{canonical}), which has the following  equilibrium
solution:
\begin{eqnarray}  \langle N_K\rangle_{\rm eq}^{\rm C}= \left [
{ {V}\over {2\pi^2}} M_{K^+}^2T  K_2 (M_{K^+}/T) \right ] \left [
{ {V}\over {2\pi^2}} M_{K^-}^2T  K_2 (M_{K^-}/T)
\right ].
\label{lim} \end{eqnarray}

\begin{figure}[htb]
\begin{minipage}[t]{70mm}
{\vskip -7.9cm
\includegraphics[width=16.5pc, height=20.2pc]{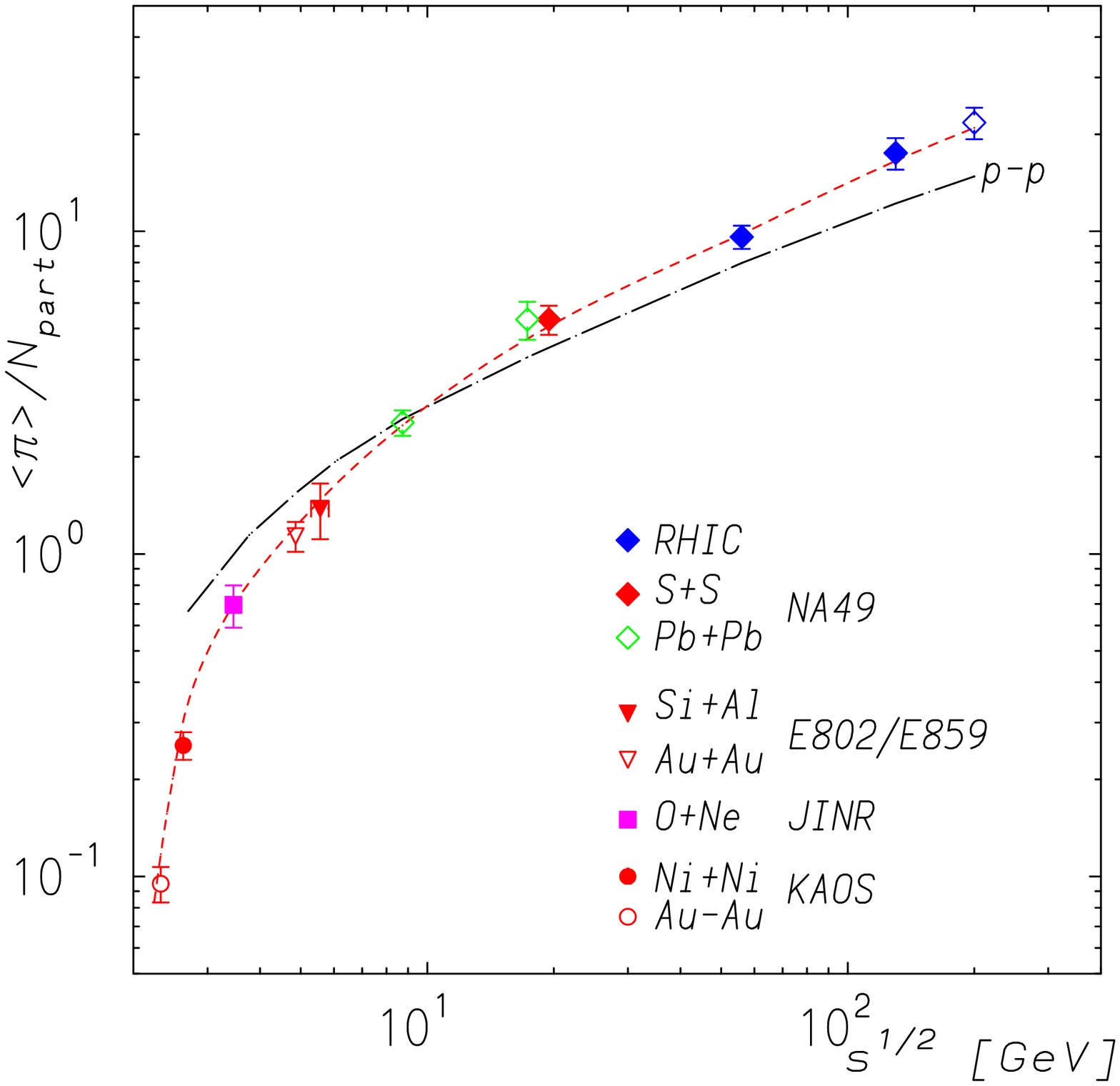}}\\
\end{minipage}
\hspace{\fill}
\begin{minipage}[t]{80mm}
\includegraphics[width=24.0pc, height=16.3pc]{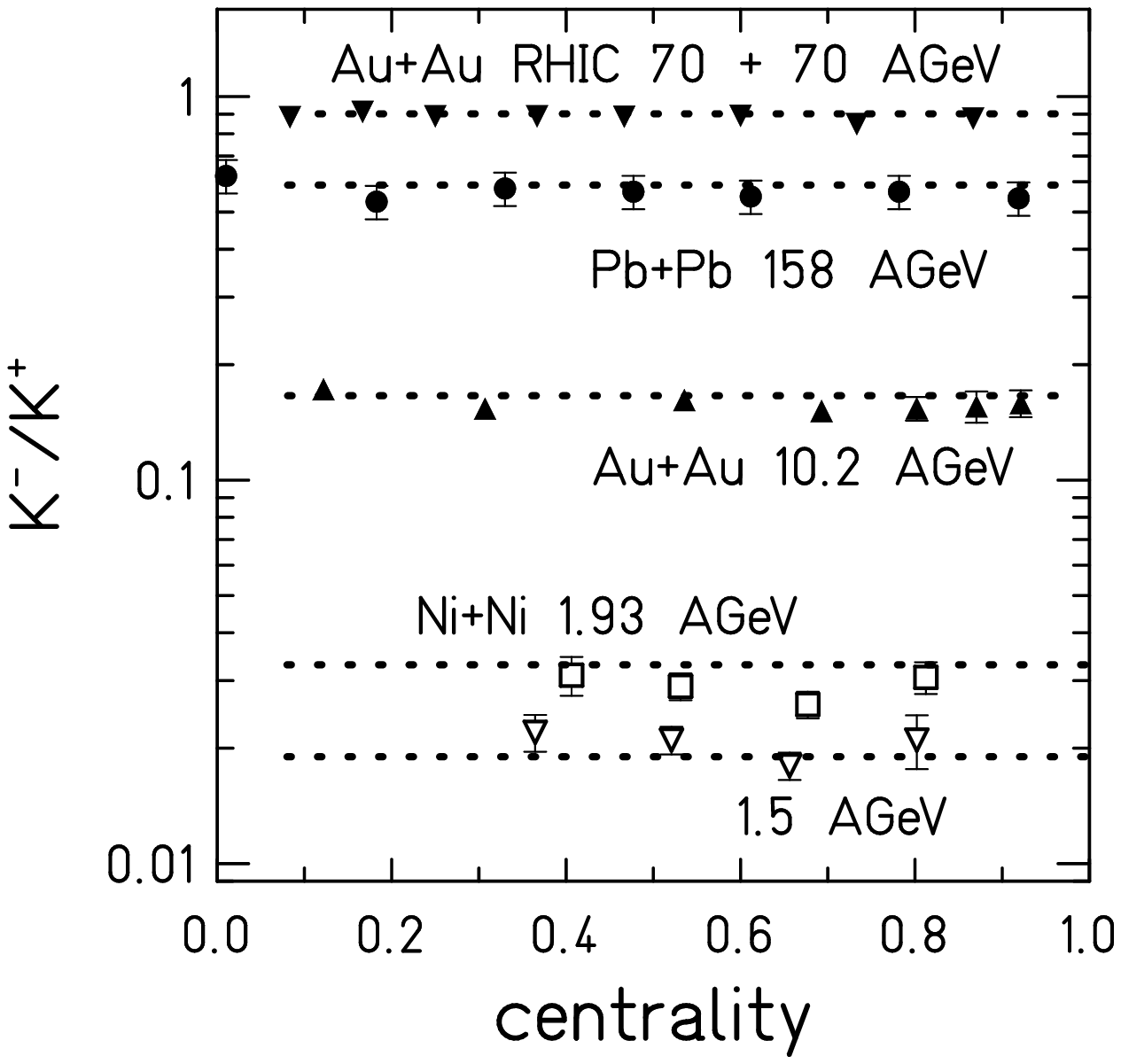}\\
\end{minipage}
\begin{minipage}[t]{169mm}
 {\vskip -1.4 true cm \caption{
The left-hand figure shows the total number of pions  per
participant ($\langle\pi\rangle /N_{\rm part}$), 4$\pi$ data in
A--A and p--p collisions versus energy.  Data at lower energies in
A--A  as well as in p--p collisions are from \cite{gr}. The RHIC
results are from \cite{rhic}. The short-dashed and dashed lines
represent the fit to the data. The right-hand figure shows the
centrality dependence of the $K^+/K^-$ ratio. Data are from the
STAR, NA49, E866, and KaoS Collaborations. The broken lines are
statistical model results.
 }}
\end{minipage}
\end{figure}

The above equation demonstrates the locality of strangeness
conservation. With each $K^+$ the  $K^-$ is produced in the same
event in order to conserve strangeness exactly and locally.
Comparing Eqs.~(\ref{15}) and (\ref{lim}), we first find  that,
for  $\langle N_K\rangle\sim 1$,  the equilibrium value is far
smaller than what is expected in the opposite limit. We also note
that the volume dependence differs in the two cases . The {\it
particle density} in the $\langle N_K \rangle \gg 1$ limit is
independent from $V$ whereas in the opposite  limit the density
scales linearly with $V$.\footnote{ In the application of the
statistical model to particle production in heavy ion collisions,
the volume of the fireball scales with the number of participating
nucleons $N_{\rm part}$. Thus, in terms of the canonical model,
depending on the abundances of produced U(1) charged particles,
one obviously expects   different centrality dependences.}

The results of  Eqs.~(\ref{15}) and (\ref{lim}) correspond to two
limiting cases of  asymptotically  large and small kaon
multiplicity in heavy ion events. In order to find the equilibrium
solution  valid for any arbitrary number of kaons per events one
needs to formulate a kinetic equation for the probability instead
of the multiplicity of produced particles \cite{ko}.

Let $P_n(t)$ ($0\leq n\le \infty$) be  the probability function
for the production of $n$,   $K^+K^-$ pairs. The probability $P_n$
tends to increase in time  owing to the transition from $n-1$ and
$n+1$ states to  $n$. It also tends to decrease since the state
$n$ makes a transition to $n+1$ and $n-1$. With  the transition
probability per unit time from  $n\to n+1$  given by
$GN_{\pi^+}N_{\pi^-}V^{-1}$ and from $n-1\to n$ described as
$LV^{-1}$, one can formulate  the general iterative master
equation for the probability  function as \cite{ko}:
\ber
\frac{dP_n}{d\tau}&=&{G\over V} N_{\pi^+} N_{\pi^-} (P_{n-1}-P_n)
\nonumber \\
&-& \frac {L}{V} \left [ n^2 P_n - (n+1)^2 P_{n+1} \right ]
\label{generaln}. \eer
The above equation is equivalent to the general rate equation
described by  Eq.~(\ref{general2}). However, contrary to
Eq.~(\ref{general2}) it can be solved for the equilibrium limit
corresponding to  an arbitrary number of produced kaons. Indeed,
converting first the above equations for $P_n$'s into a partial
differential equation for the  generating function
$g(x,\tau) = \sum_{n=0}^\infty x^n P_n (\tau)$ 
one obtains

\ber \frac {\partial g(x,\tau)}{\partial \tau} = \frac
{L}{V} (1-x) \left (x g''+ g'- \epsilon g \right ), \eer where $g'
\equiv
\partial g / \partial x$ and $\sqrt \epsilon \equiv \langle N_K\rangle_{\rm eq}^{GC}$ given by
Eq.~(\ref{15}).
The equilibrium solution for  $g_{\rm eq}(x)$ is obtained as

\ber g_{\rm eq}(x)
= \frac {1}{I_0 ( 2\sqrt \epsilon)} I_0 ( 2\sqrt {\epsilon x} ),
\label{26} \eer
where the normalization is fixed by $g(1) = \sum P_n = 1$.

The equilibrium value for the probability function $P_n$ is
expressed by
\ber P_{n,\rm eq}=\frac {\epsilon^n}{I_0 ( 2\sqrt \epsilon )
(n!)^2}, \eer
which converts to a Poisson distribution only in the limit of the
large argument  of the Bessel function.

The result for the average number of kaon pairs $K^+K^-$
in equilibrium is obtained from     $g'(1)$      and reads:
\ber \langle N_K \rangle_{\rm eq}^C= { {V}\over {2\pi^2}} m_K^2 T K_2
(M_K/T)\times
\frac {I_1 [ 2{ {V}\over {2\pi^2}} m_K^2 T K_2 (M_K/T)]}{I_0 [2{
{V}\over {2\pi^2}} m_K^2 T K_2 (M_K/T)]}. \label{neq} \eer

The above equation is a general equilibrium solution, which is
valid  for an arbitrary value of $\langle N_K\rangle$ and
obviously reproduces the asymptotic results described by
Eqs.~(\ref{15}) and (\ref{lim}). This can be seen in the most
transparent way when comparing two limiting situations: the
large--and--small $x$ (where $x$ is the argument of the Bessel
function) limit of the above equation.
%


\begin{figure}[htb]
\begin{minipage}[t]{80mm}
{
\includegraphics[width=17.5pc, height=19.3pc]{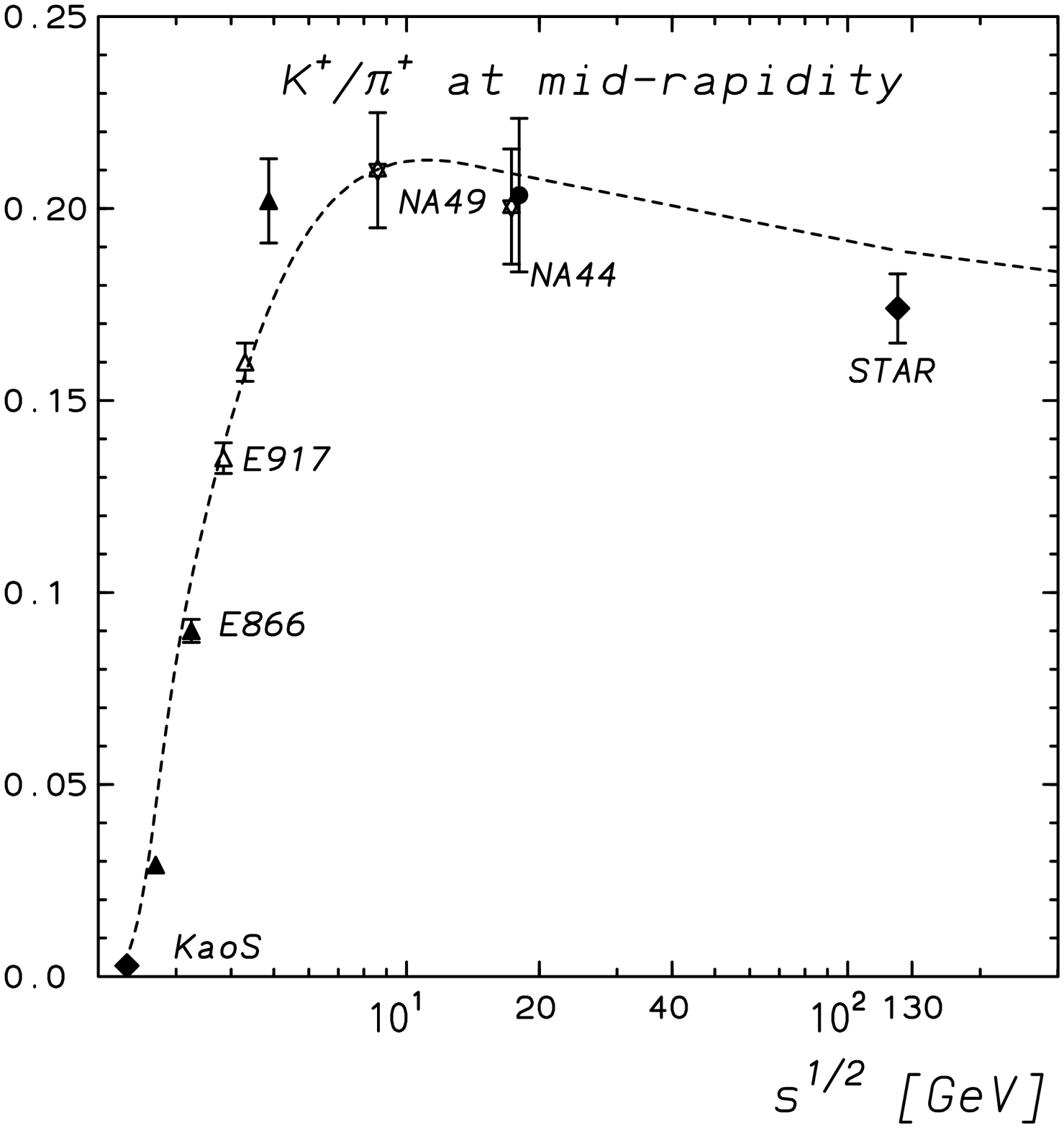}}\\
\end{minipage}
\hspace{\fill}
\begin{minipage}[t]{80mm}
\includegraphics[width=17.5pc, height=19.2pc]{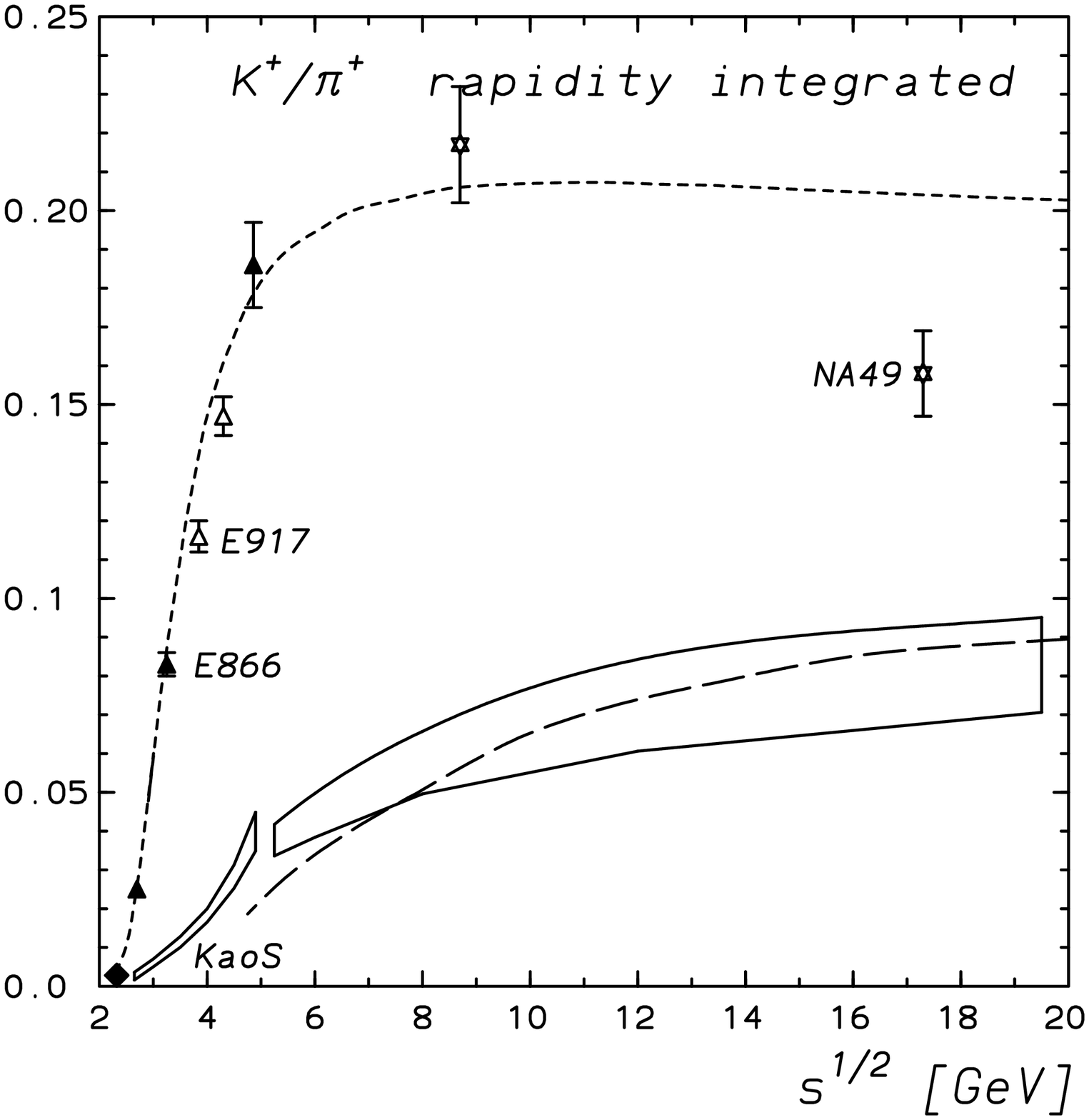}\\
\end{minipage}
 \begin{minipage}[t]{169mm}
 {\vskip -1.2 true cm \caption{
 Ratio of kaon to pion measured in heavy ion collision at different
 collisions energies. The left-hand figure corresponds to midrapidity
 data, whereas the right one represents the ratios, of fully
 integrated results. Data at SIS, AGS, SPS and RHIC are taken from
 \cite{blume,harris}. The short-dashed line  describes the results calculated
 following the freeze--out curve shown in Fig.~5. The parameterization
 of the
 p--p data from \cite{ogilvie} is indicated by the fulls-line.
 }}
\end{minipage}
\end{figure}


The equilibrium density corresponding to a large $N_K$ limit and
described by Eq.~(\ref{15}) is a standard result for the particle
density that can be obtained from the GC partition function
introduced in Eq.~(3). The general results described by Eq.~(16)
obviously require a different definition of the partition
function, which takes into account the exact conservation of
quantum numbers. This is the canonical partition function with
respect to the charge conservation. In the C approach there is no
more chemical potential under the trace as in Eq.~(3) but,
instead, the partition function is calculated by summing only
those states that  are carrying  exactly the quantum number $Q$,
that is

\begin{equation}
Z_Q^{C}(T,V)\equiv {\rm Tr}_Q[e^{-\beta H }].
\end{equation}



%


Following    Eq.~(\ref{5}) it is clear  that C and GC limits are
essentially determined by the size of $\langle \delta N^2
\rangle$, the event-by-event fluctuations of the number of
particles carrying U(1) charge. The grand canonical results
correspond to small fluctuations, i.e. $\langle \delta N^2
\rangle/\langle N \rangle^2 \ll 1$, while the canonical
description is necessary in the opposite limit.

The major difference between the canonical     and the grand
canonical treatment of the conservation laws appears through a
different volume dependence of  particle densities as well as a
strong suppression of the thermal particle phase space in the
former. The relevant parameter to measure the suppression of
particle multiplicities from their grand canonical result is seen,
in Eq.~(16), to be determined by the ratio of the Bessel functions
$ {{I_1(x)}/{I_0(x)}}$. For multistrange particles this
suppression factor has a more complicated structure
\cite{hamieh,new}.
\begin{figure}[htb]
\begin{minipage}[t]{80mm}
\includegraphics[width=17.5pc, height=20.2pc]{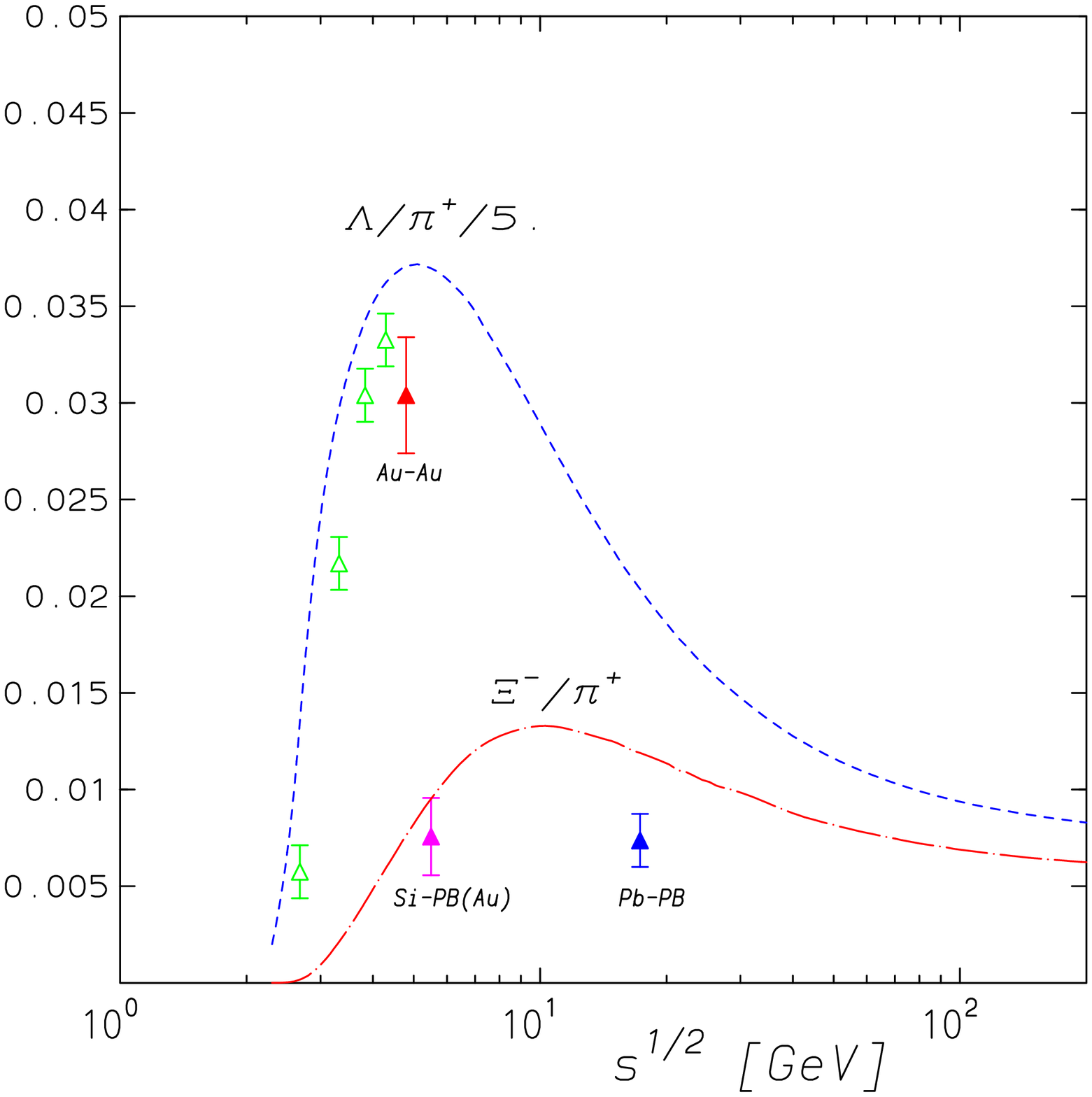}\\
\end{minipage}
\hspace{\fill}
\begin{minipage}[t]{80mm}
\includegraphics[width=17.5pc, height=19.2pc]{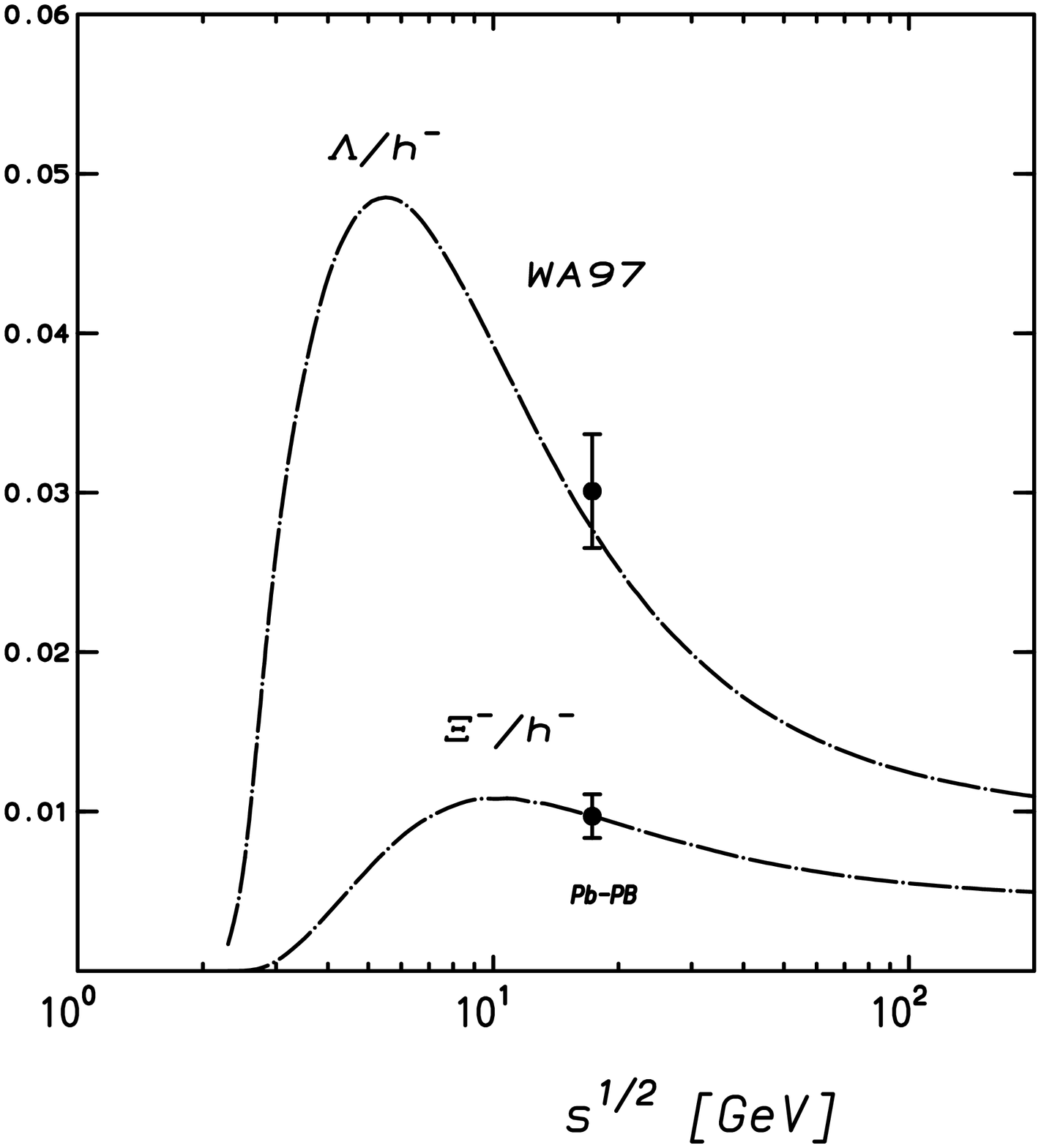}\\
\end{minipage}
\begin{minipage}[t]{169mm}
{\vskip -1. true cm \caption{ Particle ratios in A--A versus
energy. Data at the SPS are fully integrated NA49 results
(left-hand figure). The right-hand figure shows WA97 data at
midrapidity. The corresponding ratio at the top AGS was obtained
from E810 results on $\Xi^-$ measured in Si--Pb collisions in the
rapidity interval $1.4<y<2.9$ \cite{e810}, normalized to the full
phase space values of $\pi^+$ and $K^-$ yield obtained in Si--Au
collisions by E802 \cite{e866}. The lines represent statistical
model results along the unified freeze--out curve from Fig.~5.
 }}
\end{minipage}
\end{figure}

In  nucleus--nucleus collisions  the absolute values of  baryon
number, electric charge and strangeness are fixed by the initial
conditions. Modelling particle production in statistical
thermodynamics would in general require the canonical formulation
of all these quantum numbers. A detailed analysis
\cite{Hag71,marekg}, however, has shown that in heavy ion
collisions only strangeness should be treated exactly, whereas the
conservation of baryon and electric charges can be described by
the appropriate chemical potentials in the grand canonical
ensemble.

In heavy ion collisions the number of produced strange particles
depends  on the collision energy and centrality of these
collisions. At low collision energies,  at SIS/GSI for example,
the average number of  strange particles produced in an event is
much smaller than 1. Thus, here we are in the asymptotic regime of
canonical ensemble.  Figure 3a  shows the experimental data on
$K^+$ yield per participant $A_{\rm part}$ as a function of
$A_{\rm part}$ measured in Au--Au collisions at $E_{\rm lab} \sim
1$ A/GeV \cite{data}. The data are compared with the results of
the canonical statistical model shown by the dashed--line.  The
thermal parameters, the temperature and the baryon chemical
potential were chosen in such a way  as to reproduce  measured
particle multiplicity ratios of strangeness neutral particles
\cite{Cle99}. The volume parameter in the statistical operator is
assumed to scale with the number of participants. The results in
Fig.~3a clearly indicate that both the magnitude of the yield and
the strong, almost quadratic, dependence of the kaon yield on the
number of participants is well reproduced by the canonical model.

The importance of the canonical treatment of strangeness
conservation has been  shown  also at higher collision energies,
e.g.,  at the SPS or even RHIC, when considering the centrality
dependence of multistrange baryons \cite{hamieh}. In very
peripheral collisions the yield of strange particles is so
\begin{figure}[htb]
\begin{minipage}[t]{170mm}
{\hskip 2cm
{\includegraphics[width=21.5pc, height=25.5pc,angle=90]{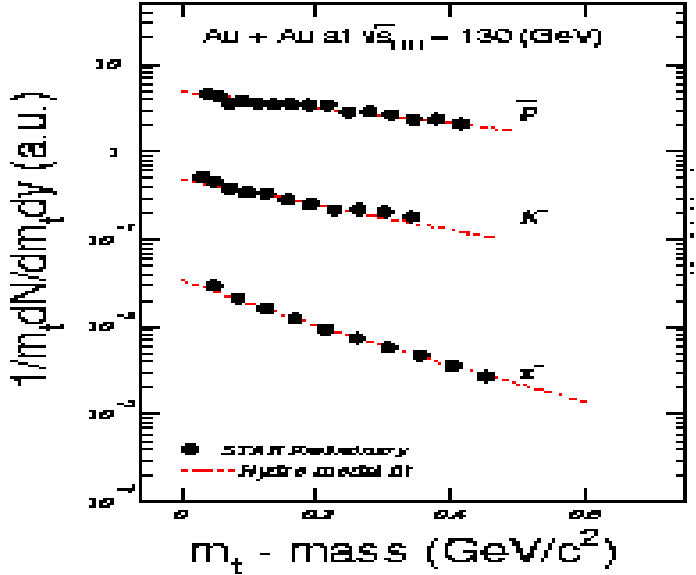}}}\\
{\vskip -1cm \caption{Transverse momentum distribution of pion,
proton and kaon obtained by STAR at $\sqrt s =~130$ GeV
\cite{nu}}.}
\end{minipage}
\end{figure}
small that  the canonical description should be applied there as
well. The canonical suppression of the thermal particle
phase--space was found to increase with the strangeness content of
the particle.

%
%

 Figure 3b  shows the multiplicity/participant of
$\Omega,~ \Xi ,$ and $\Lambda$   relative to its value in p--p or
p--A collisions \cite{hamieh}. Thermal parameters $T=145$ MeV and
$\mu_B=370$ MeV were used here and assumed  to be
centrality--independent. These values are expected in Pb--Pb
collisions at   40 AGeV. Figure 3b indicates that the statistical
model in the C ensemble reproduces the basic features of the WA97
data \cite{wa97} shown in Fig.~(4): the enhancement pattern and
enhancement saturation for large $A_{\rm part}$. Figure 4 also
demonstrates a different $A_{\rm part}$ dependence of strange and
multistrange baryons as well as a much larger enhancement at 40 A
GeV than seen in WA97 data at  top SPS. The basic predictions of
the canonical statistical model is that strangeness enhancement
from p--p to A--A collisions  should increase with decreasing
energy. This result is in contrast with UrQMD finding
\cite{bleicher} and with previous qualitative predictions that the
strangeness enhancement   is being entirely due to quark--gluon
plasma formation \cite{rafelski}.

The quantitative comparison of the canonical model with the
experimental data of WA97 has been discussed in \cite{hamieh}. The
most recent results of NA57 \cite{na57}, showing an abrupt change
of the enhancement for $\bar\Xi$ as seen in Fig.~4 are, however,
very unlikely to be reproducible in terms of the canonical
approach.

\begin{figure}[htb]
\begin{minipage}[t]{170mm}
{\hskip 3.cm
\includegraphics[width=20.3pc, height=20.3pc]{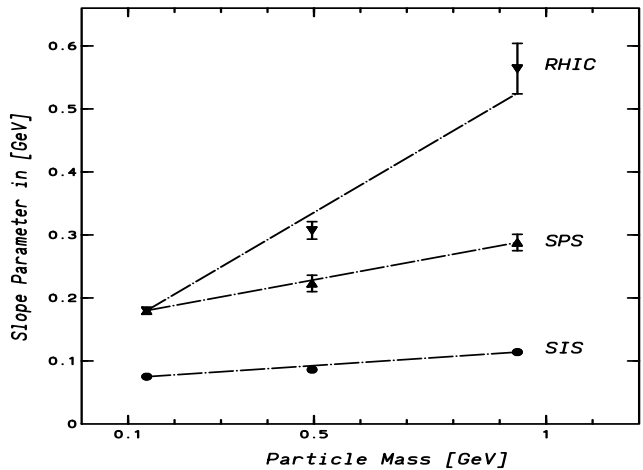}}\\
{\vskip -1.7cm \caption{ Slope parameters versus particle mass at
SIS \cite{Cle99}, top SPS \cite{hecke} and RHIC \cite{nu}.}}
\end{minipage}
\end{figure}
\section{ Particle yields and energy dependence}
In the last section  arguments were presented in favour of the
need for  that a more general treatment of U(1) charge
conservation, based on the canonical ensemble,   if one compares
the statistical model with experimental data for particle yields
in central A--A collisions at  SIS energies or even in peripheral
collisions at the SPS and RHIC. A detailed analysis of the
experimental data in heavy ion collisions from SIS through AGS has
shown that the canonical statistical model reproduces most of the
measured hadron yields.  Figure 4b  shows an example of the recent
systematic study of the   comparison of the   statistical model
with a fully integrated particle multiplicities data in central
Au--Au and Pb--Pb collisions at beam momenta of 1.7 A/GeV, 11.6
A/GeV (Au-AU) and 158 A/GeV (Pb-PB) \cite{becattini2}. The overall
agreement is seen to be very good.

Figure 5 shows the compilation of chemical freeze--out parameters
required to reproduce measured particle yields at SIS, AGS, SPS
and RHIC energies. The GSI/SIS results have the lowest freeze--out
temperature and the highest baryon chemical potential. As the beam
energy is increased a clear shift towards higher $T$ and lower
$\mu_B$ occurs. There is a common feature to all these points,
namely that the average energy per hadron is approximately 1 GeV.
{\it Chemical freeze--out} in A--A collisions is thus reached {\it
when the energy per particle drops below 1 GeV} at all collision
energies \cite{CLK}. The above phenomenological freeze--out
condition provides the relation between temperature and chemical
potential at all collision energies. This  together with the
measured ratio of pion/participant  shown in Fig.~6a establishes
the energy dependence of the two independent  thermal parameters,
the temperature and baryon chemical potential. Consequently the
definite predictions of particle excitation functions can be given
in terms of this model.  Figures 6-8 are showing statistical model
results for different particle multiplicity ratios along the
unified freeze--out curve  in comparison with experimental data.
\begin{figure}[htb]
\begin{minipage}[t]{80mm}
\includegraphics[width=17.5pc, height=18.3pc]{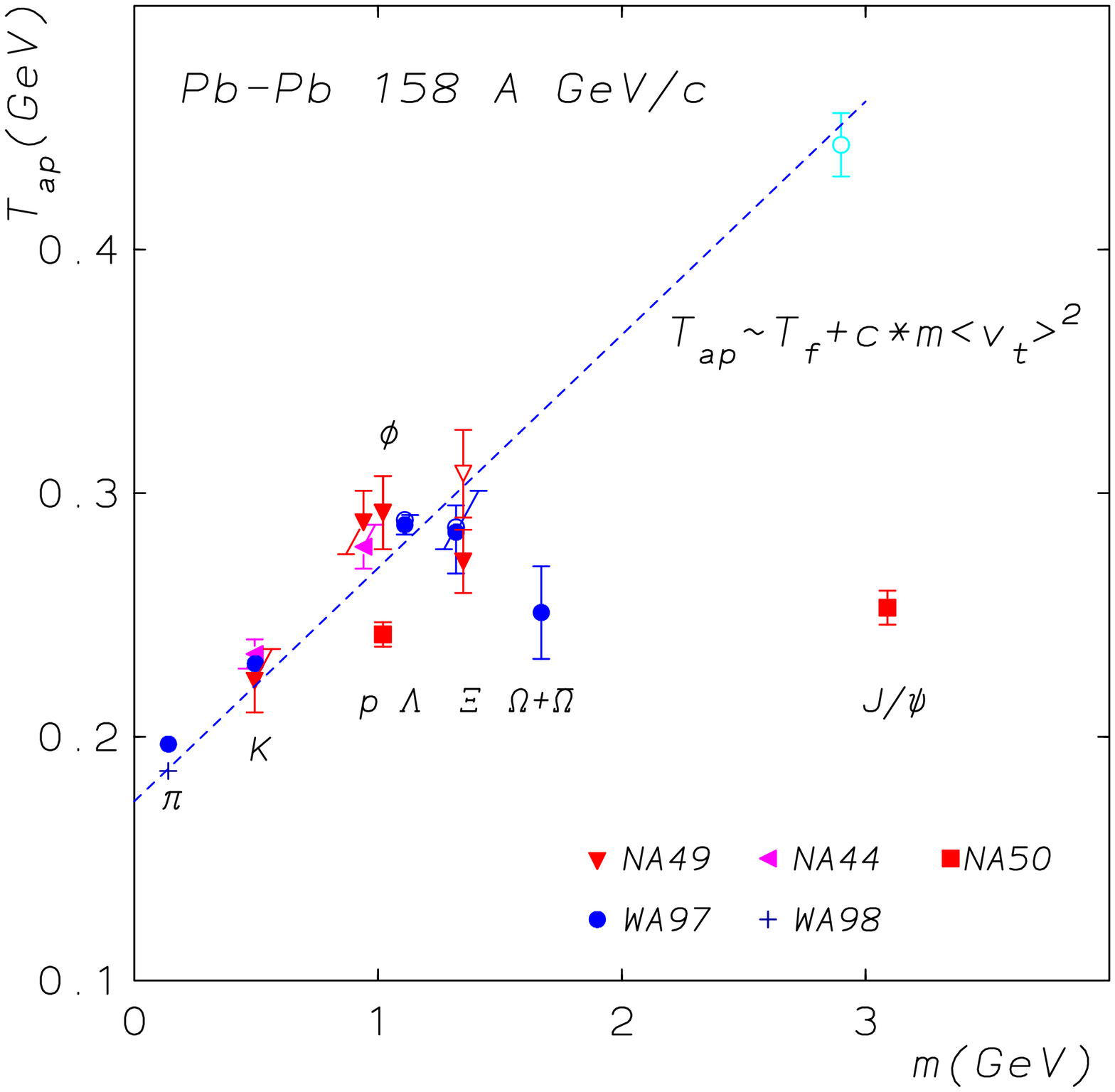}\\
\end{minipage}
\hspace{\fill}
\begin{minipage}[t]{80mm}
\includegraphics[width=17.5pc, height=18.2pc]{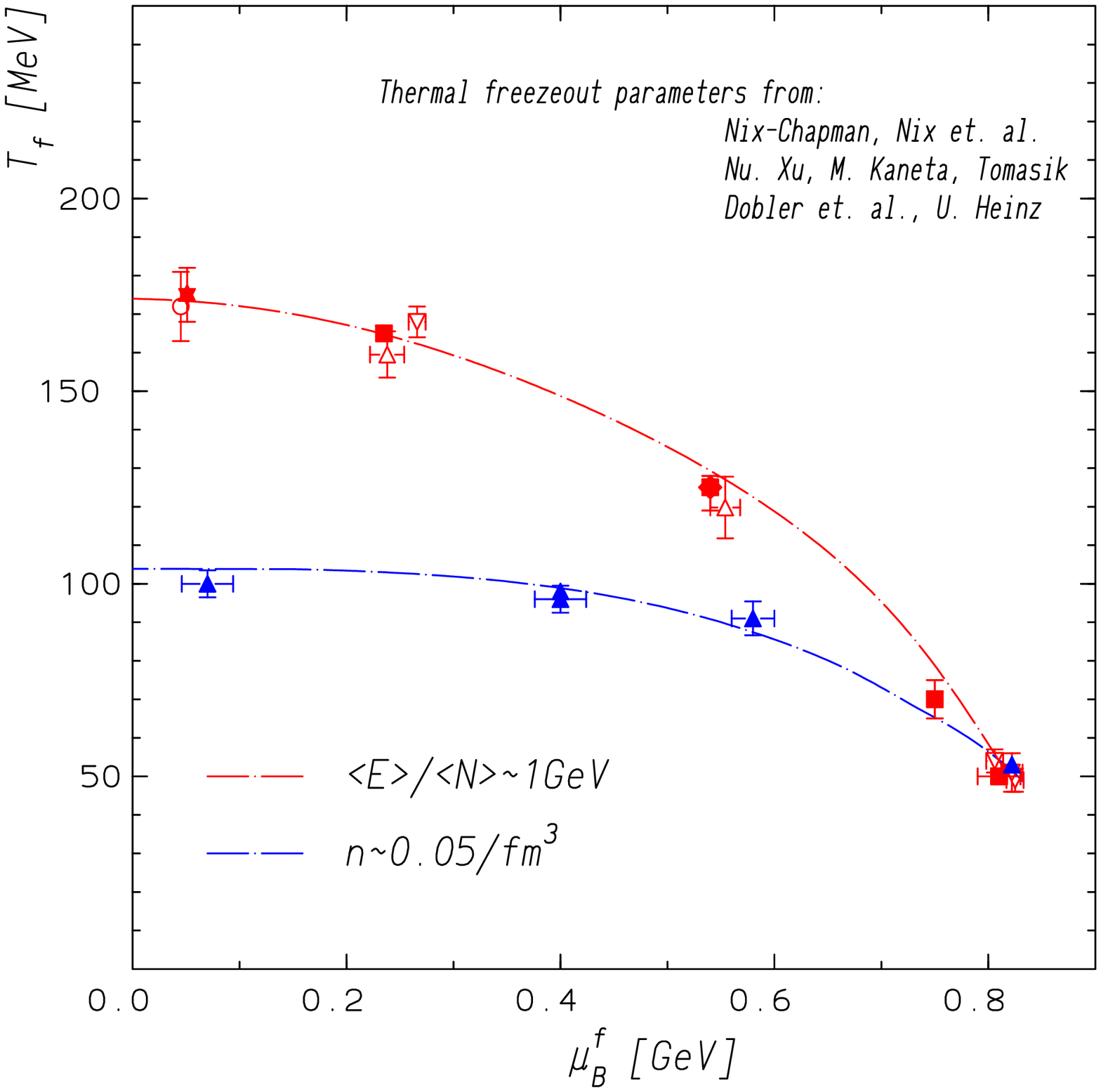}\\
\end{minipage}
\begin{minipage}[t]{169mm}
\vskip -1cm \caption{ 
 {Left: slope parameters at the SPS for different
particle species.
 For the compilation of data, see for instance  \cite{hecke,rp}. Right: Chemical and thermal freeze--out curve
from SIS to RHIC. Thermal freeze--out points are from
\cite{tomasik,rhics}}.}
\end{minipage}
\end{figure}

The statistical model predicts  that the  particle/antiparticle
ratio should be independent from centrality for all collision
energies. Dynamically this is a rather surprising result as
particles and their anti--particle   are generally  produced and
absorbed in surrounding nuclear medium in different ways. Figure
7b represents the energy and centrality dependence of the
$K^+/K^-$ ratio from SIS to RHIC. The statistical model
predictions are seen in Fig.~6b to agree remarkably well with the
data. The  measured $K^+/\pi^+$ ratio \cite{blume} is a very
abruptly increasing function of collision energy between SIS up to
top AGS. At higher energies it reaches  a broad maximum between 20
AGeV - 40 AGeV and gradually decreases up to RHIC energy
\cite{harris}. In the microscopic transport models \cite{tran} the
increase of the kaon yield with collision energy is qualitatively
expected as being due to a change in the production mechanism from
associated to direct kaon emission. However, the hadronic cascade
transport models do not,    until now, provide   quantitative
explanation of the experimental data in the whole energy range.
The statistical model in the C formulation, on the other hand,
provides an excellent description of $K/\pi$ {\it midrapidity}
data in the whole energy range from SIS up to RHIC, as seen in
Fig.~7a. The abrupt increase from SIS to AGS and broad maximum of
this ratio are  a consequence of the specific dependence of
thermal parameters on collision energy and canonical strangeness
suppression at SIS. In general, however, results with the
statistical model should be compared with 4$\pi$-integrated
yields, since strangeness does not have to be conserved in a
limited portion of phase space. A drop in the $K^+/\pi^+$  ratio
for 4$\pi$ yields has been reported from preliminary results of
the NA49 Collaboration  at 158 AGeV~\cite{blume} (see Fig.~7b).
This decrease is, however, not reproduced by the present
statistical model without further modifications, e.g.~by
introducing an additional parameter $\gamma_s\sim 0.7$
\cite{becattini2} or by formulating a statistical model {\it of
the early stage} \cite{gazdzicki}.

The appearance of the maximum in the relative strange/non-strange
particle multiplicity ratios already seen in $K^+/\pi^+$ is even
more pronounced for strange baryons/meson. Figure 8 shows the
energy dependence of $\Lambda /\pi^+$ and $\Xi^- /\pi^+$.  There
is a very clear pronounced maximum especially in the
$\Lambda/\pi^+$. This maximum is related with a rather strong
decrease of chemical potential coupled with an only moderate
increase in associated temperature with increasing energy. The
relative enhancement of $\Lambda$ is stronger than that of
$\Xi^-$. There is also a shift of the maximum to higher energies
for particles with increasing strangeness quantum number. This is
because the enhanced strangeness content of hadron  suppresses the
dependence of the corresponding ratio on $\mu_B$. The actual
experimental data both for  $\Lambda /\pi^+$ and $\Xi^- /\pi^+$
ratios shown in Fig.~8 are following the predictions of the
statistical model. However, as in the case of kaons, midrapidity
results (see Fig.~8b are better reproduced by the model than
$4\pi$ data (see Fig.~8a).

\section{Particle spectra and energy dependence}
The agreement of the equilibrium statistical model with
experimental data suggests that the collision fireball  is of
thermal nature. However,  this fireball is not a static object. At
chemical freeze--out the density inside a fireball is still so
high that its constituents undergo strong rescatterings. These
rescatterings result in thermodynamic pressure, which causes
collective expansion of the medium. Consequently the particle
transverse momentum distribution  is showing a slope,  which is
increasing with the rest mass of the particle. This is a typical
behaviour  expected from a system which undergoes  transverse
collective expansion. As an illustration,   Fig.~9 shows the
recent RHIC results of the STAR Collaboration for the $p_t$
distribution of pion, kaon and proton \cite{nu}.
\begin{figure}[htb]
\begin{minipage}[t]{70mm}
\includegraphics[width=18.5pc, height=15.9pc]{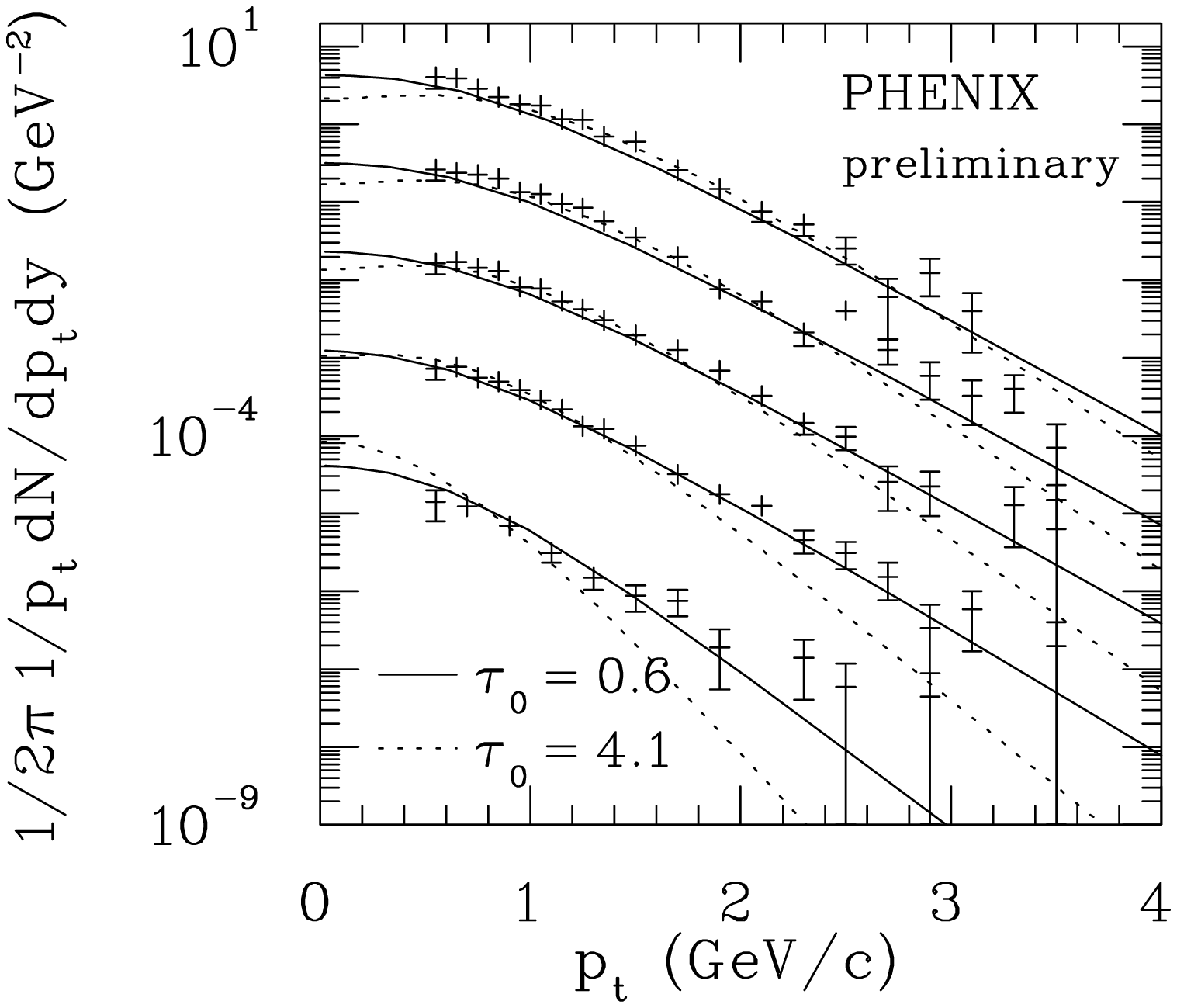}\\
\caption{ Transverse momentum distribution of antiprotons obtained
at RHIC by the PHENIX Collaboration  for different centrality
\cite{phenix}.}
\end{minipage}
\hspace{\fill}
\begin{minipage}[t]{80mm}
\includegraphics[width=18.5pc, height=15.9pc]{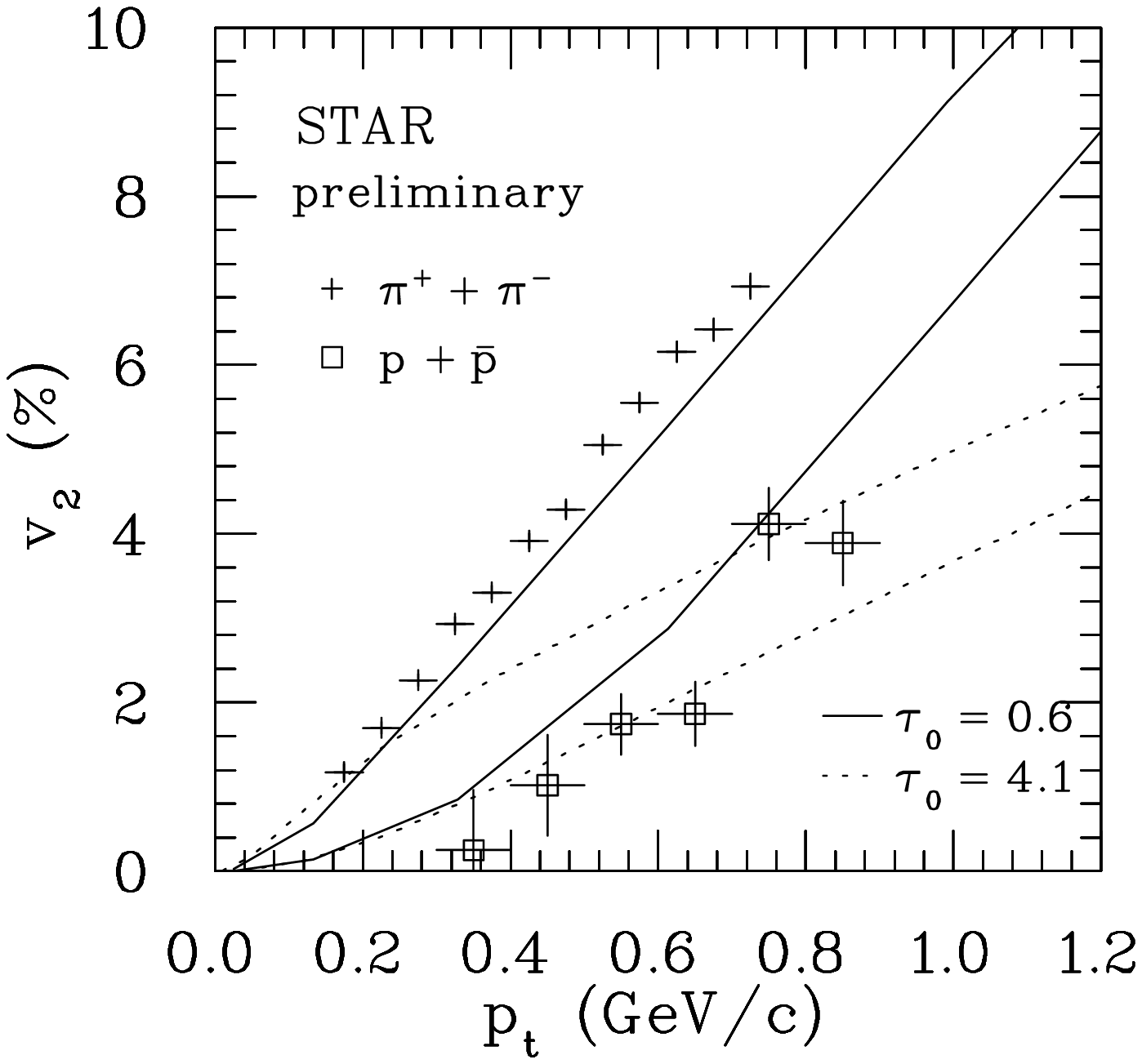}\\
{
\caption{  Elliptic flow parameter  v$_2$  of pions and
protons+antiprotons in minimum bias collisions at RHIC
\cite{phenix}.
 Lines represent
hydrodynamical model calculations from \cite{passi}.}}
\end{minipage}
\end{figure}

The collective transverse flow is, however, not only seen  in high
energy data. In Fig.~10 the dependence of the slope parameters of
pion, kaon and proton is seen to be an increasing function of
particle mass from SIS through SPS up to  RHIC energy. The
measured particle spectra at  all energies seem to be affected by
both thermal and transverse collective motion. If all particles
kinetically decoupled approximately at the same  time, then
hadronic $m_t$ spectra could be characterized by only two
parameters: average thermal freeze--out temperature $\langle
T_f\rangle$ and average flow velocity $\langle { v_t}\rangle$
\cite{wideman}. A detailed analysis has shown that this works
indeed within currently  available data and at all energies with
the exception of $\Omega$ and $J/\psi$ at the SPS  which are
showing  slopes lower than expected from flow systematics (see
Fig.~11a. These heavy particles are most likely  decoupled from
the system already at chemical freeze--out because of  their very
low rescattering cross section with surrounding hadrons. The
compilation of the thermal freeze--out parameters extracted from a
detailed analysis of experimental data
 at SIS \cite{Cle99}, the AGS \cite{tomasik}, the SPS \cite{tomasik} and most
recently at RHIC \cite{rhics} is shown in Fig.~11b. At SIS, the
thermal and chemical freeze--out coincide. At higher energies the
thermal freeze--out follows the chemical one. The freeze--out
temperature is remarkably consistent at AGS, SPS and RHIC between
100--120 MeV. Hydrodynamic evolution models \cite{10,passi,sh},
however, reported recently that at RHIC, $\langle T_f\rangle\simeq
128-140 $  could thus be slightly higher than at the SPS. Particle
$p_t$ spectra at RHIC are very well described within the concept
of thermalization and hydrodynamical evolution. Figure 12 shows an
excellent agreement of the hydrodynamical model \cite{passi} with
preliminary PHENIX data \cite{phenix} for proton $p_t$ spectra
measured at different centrality.

The transverse flow also affects other observables, which are
sensitive probes of collective motion and equilibration in heavy
ion collisions. One of these observables is the elliptic flow.
This is particularly sensitive to the initial conditions and
initial rescattering between  constituents. It describes the
azimuthal momentum-space anisotropy of particle emission in
non-central heavy ion collisions. This emission is measured in the
plane transverse to the beam direction. The parameter that
characterizes the elliptic flow is the second coefficient ${\rm
v_2}$ of an azimuthal Fourier decomposition of the particle $p_t$
spectra. Figure 13 shows that minimum bias data at RHIC for ${\rm
v_2}$ are very well described by a hydrodynamical model
\cite{passi,sh} up to $p_t\simeq 2$ GeV. Deviations are seen both
for higher transverse momentum and for detailed centrality
analysis of experimental data \cite{passi,sh}. An excellent
agreement of RHIC experimental data both for particle yields and
soft $p_t$ spectra indicates an early thermalization of the QCD
medium already on the partonic level and its further
hydrodynamical evolution.

%


\section{Summary and conclusions}

We have  reviewed some  properties of the experimental data on
particle yields and their spectra measured in A--A collisions from
SIS up to RHIC energy. We have shown that at all energies
particles seem to be  produced according to the principle of
maximal entropy, showing the statistical order of the
multiplicities. Particle spectra, on the other hand, can be
satisfactorily described by introducing in addition to thermal
also transverse collective motion. A large degree of
thermalization and collectivity in experimental data is
particularly evident at RHIC and the SPS. Here chemical
freeze--out conditions are remarkably consistent with those
expected for deconfinement, and particle spectra are well
described by transverse collective flow. Until now, however, there
is no rigorous theoretical understanding  of the thermal nature of
particle production in heavy ion collisions. At RHIC and the SPS
the appearance of the QGP in the initial state could be a driving
force towards equilibration. At low collision energies the
necessary conditions for deconfinement are not satisfied;
thermalization is thus most likely to take place through
production and rescattering of hadronic constituents.
\section{Acknowledgements}
We acknowledge stimulating discussions with P. Braun-Munzinger, J.
Cleymans, V. Koch,~~ H. Oeschler, H. Satz, R. Stock and A. Tounsi.
The partial support of the Polish Committee for Scientific
Research (KBN-2P03B 03018) is also   acknowledged.


\begin{thebibliography}{1}
%
 \bibitem{satz} For a recent review, see e.g.
S.A. Bass, M. Gyulassy, H. Stocker, and W. Greiner, J. Phys. G25
R1 (1999);

H. Satz,  Rep. Prog. Phys. 63 (2000) 1511.

\bibitem{shuryak} E. V. Shuryak, Phys. Rep. 115 (1984) 151;
J. Cleymans, R.V. Gavai and E. Suhonen, Phys. Rept. 130 (1986)
217.

\bibitem{hatsuda} T. Hatsuda, these Proceedings.

\bibitem{stock}
R. Stock,  Phys. Lett. 456 (1999) 277; Prog. Part. Nucl. Phys. 42
(1999) 295.

\bibitem{stachel}
J. Stachel, Nucl. Phys.  A654 (1999) 119c.

\bibitem{uh}
 U. Heinz, Nucl. Phys.  A685 (2001) 414 and
 A661 (1999) 349.

\bibitem{jr}
 J. Letessier, and J. Rafelski,
{ Int. J.  Mod. Phys.} E {\bf 9}, 107 (2000).
\bibitem{alam} J. Alam, et al., hep-th/010802; Ann. Phys. 286 (2001) 159

\bibitem{dinesh} D. K. Srivastava, Eur. Phys. J. 10 (1999) 487.

\bibitem{rapp} R. Rapp, Phys. Rev.
 C63 (2001) 054907.

\bibitem{wideman} U. A. Wiedemann and U. Heinz, Phys. Rep. 319
(1999) 145, and references therein.

\bibitem{gery} For a recent review, see e.g., G. E. Brown and M. Rho,
hep-ph/0103102.
\bibitem{matsui}
T. Matsui and H. Satz, Phys. Lett. B178 (1986) 416.
\bibitem{knol}
H. W. Barz, H. Schulz, B. L. Friman and J. Knol, Phys. Lett. B254
(1991) 315, Nucl. Phys. A545 (1992) 259.

\bibitem{rafelski} P. Koch, B. M\"uller, and J. Rafelski, Phys. Rep. 142
(1986) 167. J. Rafelski Phys. Lett. B262 (1991) 333.

\bibitem{Bra99}  P. Braun-Munzinger, J.
Stachel, J. P. Wessels and N. Xu, Phys. Lett. B344 (1995)  43 and
 B365 (1996) 1

P. Braun-Munzinger and J. Stachel, Nucl. Phys. A606 (1996)  320.

\bibitem{CLK} J.  Cleymans and K. Redlich,
Phys. Rev. Lett. { 81} (1998)  5284  { and references therein.}

\bibitem{bra95}P.  Braun-Munzinger, I. Heppe and J.
Stachel, Phys. Lett. { B465} (1999) 15.

\bibitem{becattini2} F. Becattini, J. Cleymans, A. Keranen, E.
Suhonen and K. Redlich,  Phys. Rev. C64 (2001) 024901.
\bibitem{tomasik} B. Tomasik, U. A. Wiedemann and U. Heinz,
Nucl. Phys. A663 (2000) 753; R. Nix, Phys. Rev. C58 (1998) 2303.

\bibitem{vesa} J. Sollfrank, et al., Phys. Rev. C55 (1997) 392.

\bibitem{huovin} P. Huovinen, et al., Phys. Lett. B503 (2001) 58,
and references therein.

\bibitem{karsch} F. Karsch, E. Laermann, and A. Peikert,  Nucl. Phys. B605
(2001)579.
\bibitem{blaizot} J. P. Blaizot, E. Iancu, and A. Rebhan,
Phys. Rev. D63 (2001) 065003.


\bibitem{rob} R. Pisarski, Phys. Rev. D62 (2000) 111501.

\bibitem{bjorken}
J. D. Bjorken, Phys. Rev. D27 (1983) 140.
\bibitem{400}
M. Agarwal, et al. (WA98 Coll.), Eur. Phys. J. C18 (2001) 651.



\bibitem{sat} L. V. Gribov, et al., Phys. Rep. 100 (1983) 1;
A.H. Mueller and J. Qiu, Nucl. Phys. B268 (1986) 427.

\bibitem{kari} K. J. Eskola, K. Kajantie, P. Ruuskanen and K.
Tuominen, Nucl. Phys. B570 (2000) 379.

\bibitem{larry} L. McLerran, and R. Venugopalan,
Phys. Rev. D49 (1994) 2233 and D49 (1994) 3352; D50 (1994) 2225.


\bibitem{nardi} D. Kharzeev and M. Nardi, Phys. Lett. B507
(2001) 121.
\bibitem{rolf}
R. Baier, A.H. Mueller, D. Schiff, and D.T. Son, Phys. Lett. B502
(2001) 51.

\bibitem{pc} K. Geiger and D. K. Srivastava, Nucl. Phys A661
(1999) 592, nucl-th/9808042;

\bibitem{tc} D. Rischke, M. I. Gorenstein, H. Stocker and W.
Greiner, Z. Phys. C51 (1990) 485;~~ G. Yen, et al., Phys. Rev. C51
(1997) 2210.



\bibitem{damian}  P. Braun-Munzinger, D. Magestro, K. Redlich, and
J. Stachel,  Phys. Lett. B518 (2001) 41.

\bibitem{Cle99} J.  Cleymans and K. Redlich,
Phys. Rev. { C60} (1999) 054908; J. Cleymans, H. Oeschler and K.
Redlich, Phys. Rev. { C59} (1999) 1663; Phys. Lett. { B485} (2001)
27.
\bibitem{cise} D. Zschiesche et al., Nucl. Phys. A681 (2001) 34.

\bibitem{b1}
F. Becattini,  Z. Phys. C69 (1996) 485; F. Becattini and U. Heinz,
Z. Phys. C76 (1997)  269.

\bibitem{kinetics}
R. Rapp and E. Shuryak, Phys. Rev. Lett. 86 (2001) 2980; C.
Greiner and S. Leupold, J. Phys. G27 (2001) L95.

\bibitem{Hag71}
R. Hagedorn, Thermodynamics of stron interactions, CERN Report
71-12 (1971); E.V. Shuryak, Phys. Lett. { B42} (1972) 357;  K.
Redlich and L. Turko, Z. Phys.  { B97} (1980) 279; R. Hagedorn and
K. Redlich, Z. Phys. { A27} (1985)  541.
\bibitem{ko}  C.M. Ko, V. Koch, Z. Lin, K. Redlich,
M. Stepanov and X.N. Wang, nucl-th/0010004, Phys. Rev. Lett. 86
(2001) 5438.
\bibitem{hamieh}
J. S. Hamieh, K. Redlich and A. Tounsi, Phys. Lett. { B486} (2000)
61.

\bibitem{koch} T. Matsui, B. Svetitsky and L.D.
McLerran, Phys. Rev. { D34}  (1986) 783.


\bibitem{data} A. Wagner et al. (KaoS Coll.), Phys. Lett. B420 (1998)
20, C. Muntz et al. (KaoS Coll.) Z. Phys.  C357 39.
\bibitem{new} P. Braun-Munzinger, J. Cleymans, H. Oeschler and K.
Redlich, nucl-ph/0106066, Nucl. Phys. A in print.

\bibitem{wa97} E. Andersen, et al. (WA97 Coll.),
Phys. Lett. B449 (1999) 401.
\bibitem{marekg} M. Gazdzicki and M. Gorenstein, Phys. Lett. B483 (2000) 60.

\bibitem{bleicher} M. Bleicher, W. Greiner, H. St\"ocker, and N. Xu,
Phys. Rev. C62 (2000) 061901.
\bibitem{na57} N. Carrer (NA57 Coll.), in
Proceedings of QM2001.

\bibitem{blume} Ch. Blume (NA49 Coll.), in Proceedings of
 QM2001.

\bibitem{harris} J. Harris (STAR Coll.), in Proceedings of QM2001.
\bibitem{tran}
 W. Cassing, Nucl. Phys. A661 (1999) 468c.

\bibitem{ogilvie} C. A. Ogilvie, nucl-ex/010410, J. C. Dunlop and C. A. Ogilvie, Phys. Rev. C61 (2000) 031901.
\bibitem{gazdzicki} M. Gazdzicki,  et al., Z. Phys. C65 (1995) 215,
Acta Phys. Pol.  B30 (1999) 2705.


\bibitem{e810} S. E. Eiseman et al. (E810 Coll.), Phys. Lett. B297
(1992) 44 and  B325 (1994) 322.

\bibitem{e866} T. Abbott et al., E802 Coll., Phys. Rev. C60 (1999)
044904; Y. Akiba et al. (E802 Coll.), Nucl. Phys. A590 (1995)
179c.

\bibitem{nu} Nu Xu, nucl-ex/0104021,
in Proceedings of QM2001.

\bibitem{rp} K. Redlich, hep-ph/0105104.

\bibitem{hecke} H. van Hecke, H. Sorge and N. Xu, Nucl. Phys. A661
(1999) 493c.

\bibitem{rhics} J. Burward-Hoy (PHENIX Coll.),
presented at {\it Workshop on Thermalization in Heavy Ion
Collisions}, BNL, July 2001.

\bibitem{10} U. Heinz, hep-ph/0109006.

\bibitem{passi} P. Huovinen, hep-th/0108033 and these Proceedings.
\bibitem{sh}
D. Teaney, J. Lauret and E. Shuryak, Phys. Rev. Lett 86 (2001)
4783; nucl-th/0104041.

\bibitem{phenix} J. Velkovska
(PHENIX Coll.), nucl-ex/0105012; C. Adler et al. (STAR Coll.),
nucl-ex/0107003.

\bibitem{gr} M. Gazdzicki, and D. Roehrich, Z. Phys. C66
(1995) 77.

\bibitem{rhic} The 4$\pi$ results at $\sqrt s=56$ GeV are estimated from
PHOBOS mid-rapidity data from \cite{phobos} scale by the same
factor as required in p--p collisions when going from midrapidity
to full phase space. The results at $\sqrt s=130$ GeV are from
\cite{brahms1}. The 4$\pi$ results at $\sqrt s=200$ GeV were
estimated by multiplying the $\sqrt s=130$ GeV result by  the same
factor (modulo increase in rapidity interval) as recently measured
by PHOBOS at midrapidity \cite{phobos}.

\bibitem{brahms1}
I.G. Bearden, et. al., nucl-ex/0108016.

\bibitem{phobos}
B.B. Back, et. al., nucl-ex/0105001; Phys. Rev. Lett. 85 (2000)
3100; nucl-ex/0108009.






\end{thebibliography}
\end{document}